\documentclass{elsart}
\usepackage{color,graphics,epsfig,amssymb,amsmath,rotating}
\usepackage{lineno}

\def\ket#1{\left|\,#1\,\right>}

\def\erw#1{\left<\,#1\,\right>}

\newcommand{\depol}{D}
\begin{document}

\begin{frontmatter}
\title{The HERMES Polarized Hydrogen and Deuterium Gas Target
in the HERA Electron Storage Ring}

%%%%%%%%%%%%%%%%%%%%%%%%%%%%%%%%%%%%%%%%%%%%%%%%%%%%%%%%%%%%%%%%%%%%%

\collab{The HERMES Collaboration}

\author[18]{A.~Airapetian},
\author[30]{N.~Akopov},
\author[30]{Z.~Akopov},
\author[8,30]{M.~Amarian},
\author[16]{A.~Andrus},
\author[8]{E.C.~Aschenauer},
\author[29]{W.~Augustyniak},
\author[30]{R.~Avakian},
\author[30]{A.~Avetissian},
\author[12]{E.~Avetissian},

% ...
\address[8]{DESY, 15738 Zeuthen, Germany}
% ...
\address[12]{Istituto Nazionale di Fisica Nucleare, Laboratori Nazionali di Frascati, 00044 Frascati, Italy}
% ...
\address[16]{Department of Physics, University of Illinois, Urbana, Illinois 61801-3080, USA}
% ...
\address[18]{Randall Laboratory of Physics, University of Michigan, Ann Arbor, Michigan 48109-1120, USA }
% ...
\address[29]{Andrzej Soltan Institute for Nuclear Studies, 00-689 Warsaw, Poland}
\address[30]{Yerevan Physics Institute, 375036 Yerevan, Armenia}
% ...
\thanks[40]{Corresponding author: Tel:+49-40-89981963; e-mail: Paolo.Lenisa@desy.de}

%%%%%%%%%%%%%%%%%%%%%%%%%%%%%%%%%%%%%%%%%%%%%%%%%%%%%%%%%%%%%%%%%%%%%%%%%%
\begin{abstract}

The HERMES hydrogen and deuterium nuclear-polarized gas 
targets have been in use since 1996 with the polarized
electron beam of HERA at DESY to study the spin structure of the
nucleon.  Polarized atoms from a Stern-Gerlach Atomic Beam Source 
are injected into a storage cell internal to the HERA electron ring.
Atoms diffusing from the center of the storage cell into a side tube are 
analyzed to determine the atomic fraction and the atomic polarizations.
The atoms have a nuclear polarization, the axis of which is defined
by an external magnetic holding field. The holding field was 
longitudinal during 1996-2000, and was changed to transverse in 2001.
The design of the target is described, 
the method for analyzing the target polarization is outlined,
and the performance of the target in the various running periods is
presented. 

\end{abstract}
\end{frontmatter}

%\begin{linenumbers}

%%%%%%%%%%%%%%%%%%%%%%%%%%%%%%%%%%%%%%%%%%%%%%%%%%%%%%%%%%%%%%%%%%%%%
\section{Introduction}

The HERMES experiment at HERA (DESY, Hamburg) was designed to study the 
quark-gluon spin structure of the nucleon by deeply inelastic scattering (DIS)
of longitudinally polarized electrons from an polarized gas
target of hydrogen and deuterium atoms internal to a
storage ring\cite{TDR,Rith}. The 27.5\,GeV electron beam stored in the HERA 
storage ring with currents up to 50\,mA is
polarized by the Sokolov-Ternov effect~\cite{sokolov}, with 
a time constant of about 30 minutes.  A longitudinal spin
direction is obtained where the experiment is located within the HERA east 
straight section by means of a pair of spin rotators~\cite{barber}.
In a machine with one pair of spin rotators before the 2001 upgrade,
a typical polarization of 60\% was achieved. 

The HERMES experiment~\cite{spectrometer} was proposed in 1990, and 
is designed to 
run in parallel with the ZEUS and H1 collider experiments without 
significantly reducing the electron beam lifetime, which is of the order 
of 10 hours.
This limits the target areal density to about $10^{15}$ 
H-atoms/cm$^{2}$ 
and excludes the use of any solid target material. A target employing a cold
storage cell fed by a a polarized
atomic beam source was proposed and implemented.
Polarized gas targets for storage rings were reviewed recently by Steffens
and Haeberli~\cite{steffens}.

% During the first year of operation, 1995, the target was operated with 
% polarized $\mathrm{^{3}He}$ gas produced by a different source based on optical
% pumping~\cite{he3}. 
Since 1996, the present target apparatus for polarized hydrogen and deuterium 
has been in operation. During the years 1996-97 longitudinally
polarized hydrogen ($H_{||}$) was used, while longitudinally polarized 
deuterium was employed in 1998-2000 ($D_{||}$), and from 2002 onwards,
transversely polarized hydrogen ($H_{\perp}$) has been in use. 

Apart from a high areal density and reliable operation over many years,
the most challenging requirement for the internal target 
was to determine the effective target 
polarization seen by the beam. 
The target polarization cannot be measured by making use of scattering 
processes involving the high-energy electron beam due to the small cross
sections involved.
Nuclear vector polarization $P_z$ and, for deuterium
also the tensor polarization $P_{zz}$ were to be measured, as well as the electron
polarization $P_e$ for diagnostic purposes. 
The required accuracy was $\delta P = 0.03$ absolute
uncertainty on the average polarization over long measuring periods, and 1\,\%
statistical accuracy within a couple of minutes in order to monitor
possible changes in time. This goal was achieved by employing a sampling
technique in conjunction with detailed simulations and frequent calibration
measurements. 

In the present article, an overview of the HERMES polarized hydrogen and 
deuterium gas target is given with reference to the various papers describing
the subsystems of the target. The atomic beam source 
(ABS)~\cite{Hermes:ABS}, the storage cell~\cite{Hermes:cell}, 
the Target Gas Analyzer (TGA)~\cite{Hermes:TGA} and the 
Breit-Rabi polarimeter (BRP)~\cite{Hermes:BRP} are summarized in Section 2.
Section 3 describes the analysis applied to the raw TGA and BRP signals and
the several calibrations required to derive measured values for the
degree of dissociation and the atomic polarizations
of the sample gas\footnote{In this article, the term ``atomic polarizations" 
refers to the polarizations of the electrons and nuclei in the atoms}. 
Section 4 reviews the various physical processes
inside the storage cell that directly influence the TGA and BRP 
analyses~\cite{Hermes:recomb,Hermes:beamdepol,Hermes:beta,molflow}.
The results of the target analysis are reported in Sections 6 and 7.
Section 8 mentions a selection of measurements carried out by HERMES that
crucially depended on the polarized target.
The paper is concluded in Section 9.

%%%%%%%%%%%%%%%%%%%%%%%%%%%%%%%%%%%%%%%%%%%%%%%%%%%%%%%%%%%%%%%%%%%%%
%\input{setup_paolo.tex}
\section{Survey of experimental apparatus}

\begin{figure*}
\begin{center}
\includegraphics[width=\textwidth]{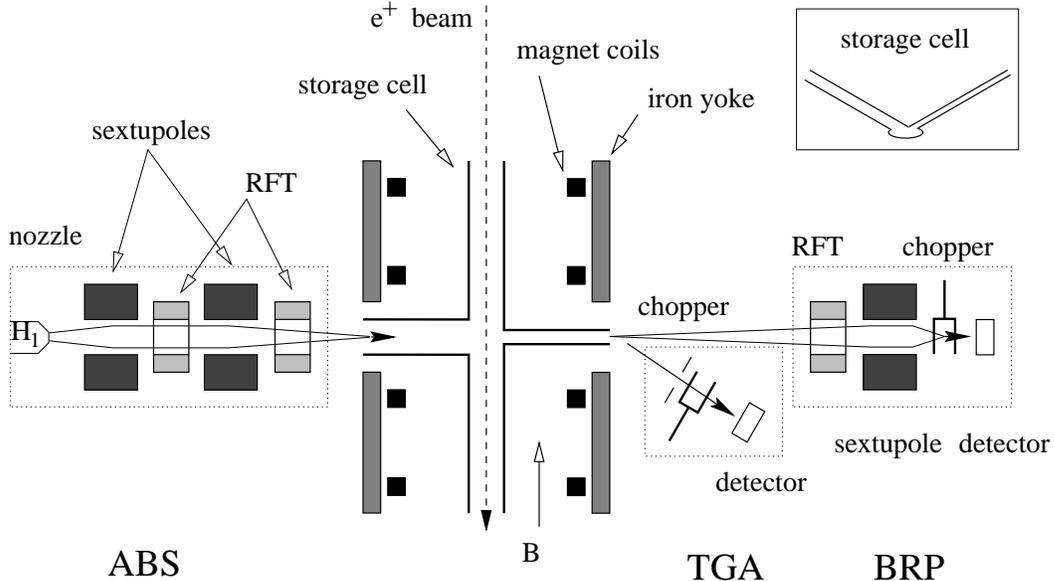}
\end{center}
\caption{Schematic view of the HERMES longitudinally 
polarized target. From left to right:
Atomic Beam Source (ABS), target chamber with cell and magnet, and diagnostic 
system composed by Target Gas Analyzer (TGA) and Breit-Rabi 
Polarimeter (BRP).  The locations of the radio-frequency transition (RFT)
units are indicated.} 
\label{fig:target}
\end{figure*}

The HERMES polarized hydrogen and deuterium target, 
schematically shown in Fig.~\ref{fig:target}, 
consists of an atomic beam source (ABS) that injects a 
spin polarized beam of H or D atoms into a thin-walled storage cell.
A sample of gas diffuses from 
the center of the cell into a
Breit Rabi Polarimeter (BRP) that measures the atomic polarizations,
or into
a Target Gas Analyzer (TGA) that measures the relative atomic and molecular 
content of the gas. A magnet surrounding the storage cell provides a holding
field defining the polarization axis and preventing spin relaxation by
effectively decoupling the magnetic moments of electrons and nucleons.
% The HERMES experiment was provided with longitudinally polarized 
% hydrogen/deuterium
% nuclei untill the end of 2000. During the HERA shutdown in 2001, 
% the target was changed to enable measurements with 
% transverse polarized hydrogen.

\subsection{Target chamber}
\begin{figure*}
\begin{center}
\includegraphics[width=\textwidth]{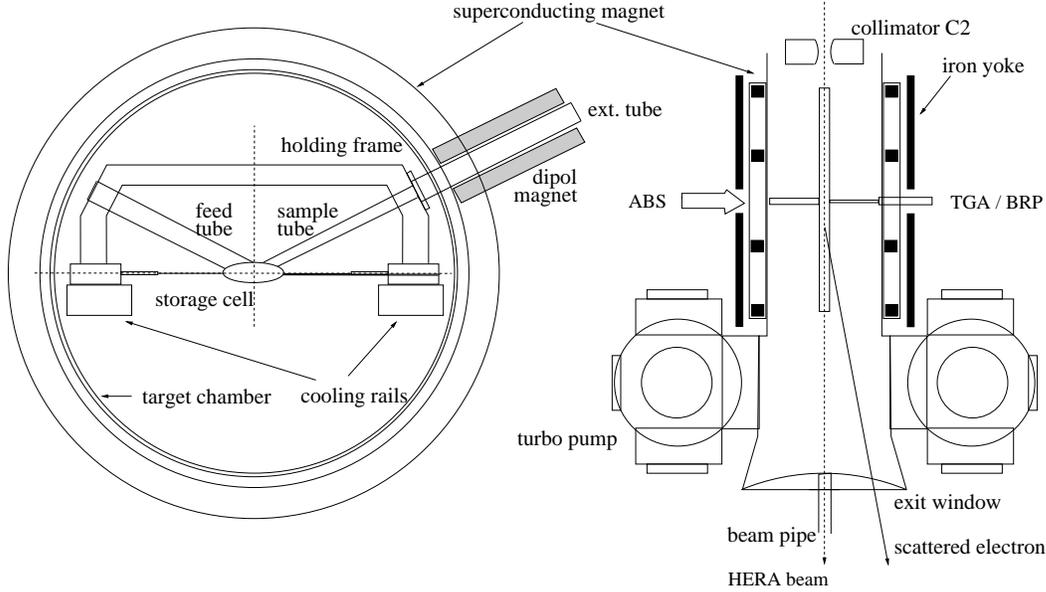} 
\end{center}
\caption{Longitudinal 
target chamber and superconducting magnet viewed from downstream with 
respect to the HERA beam direction (left), and from above (right).}
\label{fig:chamber}
\end{figure*}

The target chamber shown in Fig.~\ref{fig:chamber} is evacuated by two turbo-molecular pumps with a combined
speed of 4400\,$\mathrm{l\,s^{-1}}$. During operation the chamber pressure is
 typically in the $10^{-7}$\,mbar range due to the high gas load from
the atomic beam, which enters the chamber from $30^{\circ}$ above 
the horizontal
plane. The storage cell and its supporting cooling rails 
are cantilevered from the upstream\footnote{Upstream and downstream
mean with respect to the HERA lepton beam direction} end flange.
%Until September 2000 the HERMES target was operated with a magnetic holding
%field up to 350 mT which was longitudinally oriented with respect 
%to the HERA lepton beam. The magnet used to
%provide such a field is a superconducting magnet cooled by gaseous helium
%from a storage dewar below the target platform. The cold helium gas 
%emerging from 
%the superconducting magnet flows through the aluminum rails on
%which the storage cell is mounted. A schematic of the target vacuum chamber
%surrounded by the coils and the iron yoke of the magnet is
%shown in the lower panel of Fig.~\ref{fig:target}.
%In the same picture, the pump system needed to evacuate 
%the chamber, consisting of two
%turbo molecular pumps with a total pumping speed of 4400\,$\mathrm{l\,s^{-1}}$ 
%is also depicted.

A 0.3\,mm thick stainless steel exit window on the downstream end
of the target chamber allows the scattered electrons and 
hadronic fragments to leave the target chamber and to be detected by the
HERMES spectrometer~\cite{spectrometer}. This thin window is
integral with the following thin-walled beam tube with no intervening
flange, in order to minimize the high-$Z$ material in this region
and reduce shower production by $\pi^0$ mesons produced at small angles.
About 1\,m upstream of the target cell, an adjustable collimator
(labeled C1) is located to stop direct synchrotron radiation.  It 
normally has the smallest scaled aperture in the HERA ring, and is 
opened during electron injection~\cite{spectrometer}.
A fixed collimator (labeled C2) just upstream of the storage cell
protects it from synchrotron light scattered by C1 and from 
exposure to electron showers from C1.

%\subsection{Longitudinal magnet}
%The superconducting magnet consists of four coils around the target chamber
%as indicated in the right hand part Fig.~\ref{fig:targetchamber}. 
%It provides the longitudinal
%holding field over the storage cell of up to 350 mT defining 
For longitudinal running the storage cell and vacuum chamber are immersed
in a longitudinal field generated by a superconducting magnet. The magnet
consists of four coils and an iron yoke designed to provide optimum
field uniformity over the length of the storage cell while providing a central
gap for the atomic beam to enter the cell. The magnetic field (up to 350\,mT)
provides the quantization
axis for the spins of the polarized atoms in the storage cell, and
decouples the spins of the nucleons and electrons. 
The passing HERA
beam is bunched to provide high luminosity for the collider experiments
H1 and ZEUS.  As discussed in Sec.~\ref{sec:bunch_depol}, 
this may induce resonant nuclear depolarization of the target gas.
The induced 
resonances can be avoided by proper setting of the holding field,
provided that its homogeneity is sufficient~\cite{Hermes:beamdepol}. 
In the case of hydrogen, the
maximum permissible variation of the magnetic field along the storage cell
is $1.5\%$. For operating with deuterium gas, this requirement 
was found to be less critical~\cite{Baumgarten:thesis2}.
\begin{figure*}
\begin{center}
\begin{tabular}{c}
\includegraphics[width=0.85\textwidth]{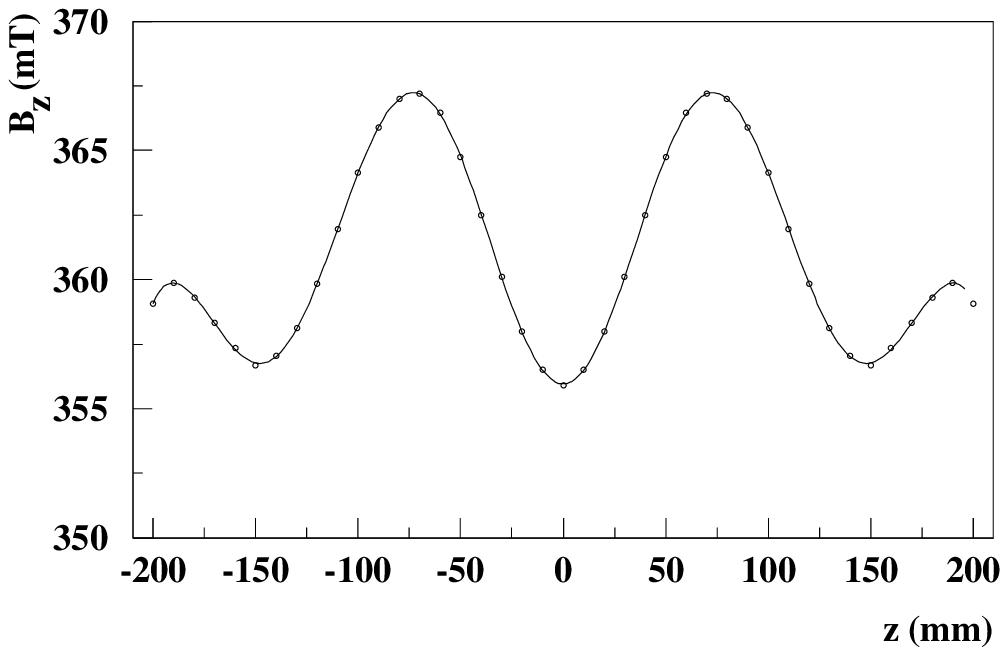}\\
\includegraphics[width=0.85\textwidth]{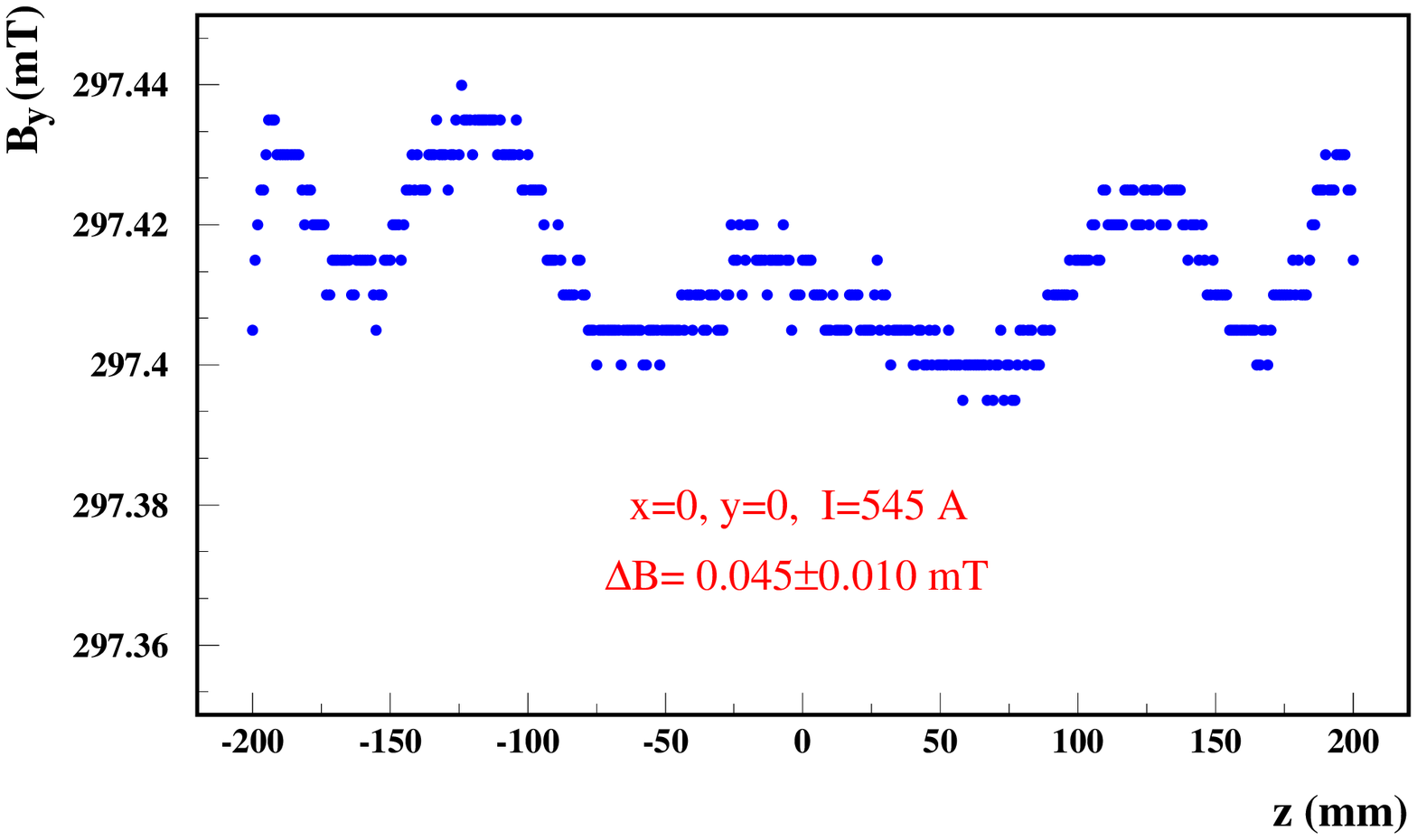}
\end{tabular}
\end{center}
\caption{Upper panel: field strength of the longitudinal target magnet 
measured 
along the HERA beam direction {\it{z}}, covering the full length of
 the storage cell. The line shows a polynomial fit to the measurement.
Lower panel:  transverse magnet field uniformity measured along the {\it{z}}
axis at the nominal field strength $B=297\,$mT. Deviations 
of $\Delta B_y=0.15\,$mT (vert.) and $\Delta B_x=0.60\,$mT (horiz.)
have been measured within the cell volume.}
\label{fig:scm_field}
\end{figure*}

\subsection{Transverse magnet}
After the end of the longitudinal running in 2000, 
the HERMES target was modified 
for measurements with a hydrogen gas target in a transversely oriented 
magnetic holding field. 
The intensity of the transverse magnetic field was limited 
by the amount
of synchrotron radiation power generated by the deflection of the beam by
the target magnet (5\,kW maximum), so it was decided to build a normal 
conducting magnet. 
%The use of a transverse field requires
%additional care in the setting to avoid bunch field induced 
%resonant depolarization by the HERA beam, 
As will be explained in Sec.~\ref{sec:bunch_depol},
a better homogeneity of the transverse target magnet of
$\Delta B \le 0.15$\,mT at a field value of about 300 mT was required.

The geometrical constraints imposed by the HERMES 
acceptance made it difficult 
to design a magnet fulfilling the uniformity requirement. After
construction, the field uniformity was measured and 
improved by shimming the pole tips. 
At a field strength of $B$=297$\,$mT, maximum deviations of 
$\Delta B_z=0.05$\,mT, $\Delta B_y=0.15$\,mT and $\Delta B_x=0.60$\,mT 
within the storage cell volume have been achieved, 
$z$ being the longitudinal direction, $y$ the vertical and $x$ 
the horizontal axis. The magnet was installed in July 2001.
The strength of the transverse field vs. z is shown in the lower plot of
Fig.~\ref{fig:scm_field}.

\subsection{The storage cell}
\label{sec:cell}
\begin{figure*}
\begin{center}
\includegraphics[width=0.75\textwidth]{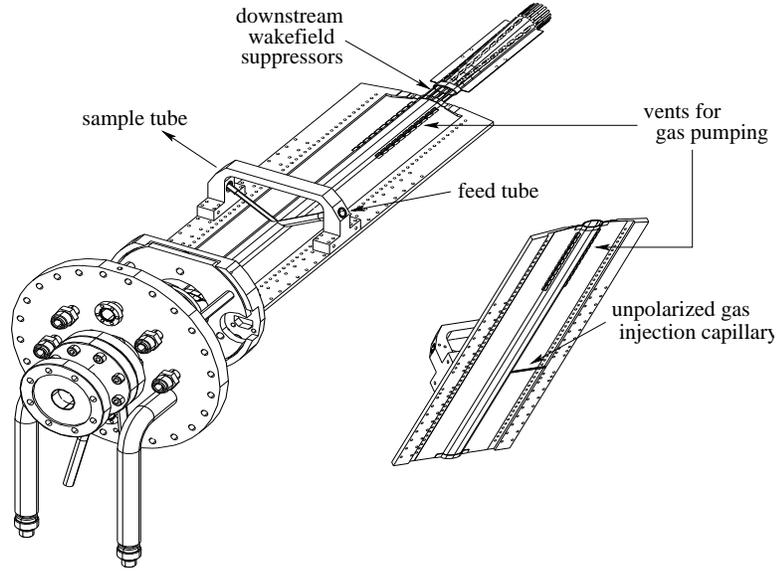} 
\end{center}
\caption{The storage cell and its support flange.}
\label{fig:cell}
\end{figure*}

The HERMES storage cell~\cite{Hermes:cell} 
is made of two 75\,$\mu m$ thin pure aluminum sheets,
which are tightly spot-welded together. It is 400\,mm long and has an elliptical
cross-sectioned shape of $21 \times 8.9\,\mathrm{mm^2}$. 
All the aluminum parts of the
cell are coated with Drifilm~\cite{Drifilm} to minimize gas depolarization and
recombination caused by wall collisions~\cite{bou63a,bou63b,bou65,and63}. 
A vented extension of the cell continues about
147\,mm downstream of the cell before joining the thicker-walled
wakefield-suppressing transition tube, to ensure that all scattered
particles inside the spectrometer detector acceptance pass through the 
minimum thickness of material (see Fig.~\ref{fig:cell}).
The wake field suppressors are made of 100\,$\mu$m thick titanium, and
confine the electromagnetic high frequency field of the bunched HERA beam,
forming a gradual electrical transition between the storage cell 
and the HERA beam pipe.

Cells with three different elliptical cross sections 
have been used thus far. The mentioned size of
 $21 \times 8.9\,\mathrm{mm^2}$ is the one in use since 
December 1999, called
the medium-sized cell. The largest was used until December 1999, and
had a cross section of  $29 \times 9.8\,\mathrm{mm^2}$. 
(A small one ($19 \times 8.0\,\mathrm{mm^2}$) was tested in January 1999
and found to be too small.)
Tab.~\ref{cellsize} summarizes dimensions and conductances of the three
cells that have been used. The conductance values apply to gas injected into
the cell center.
%%%%%%%%%%%%%%%%%%%%%%%%%%%%%%%%%%%%%%%%%%%%%%%%%%%%%%%%%%%%%%%%%%%%%%%
\begin{table}
\begin{center}
\begin{tabular}{|c|c|c|c|}
\hline
 period of use & cross section & conductance\\
 &  $a\times b$~(mm) & (l/s)\\
\hline
05.96 - 11.99 & $29.8 \times 9.8$ & 1.45$\sqrt{T/M}$ \\ \hline 
01.99 (test) & $19.8 \times 8.0$ & 0.61$\sqrt{T/M}$ \\ \hline 
12.99 - now & $21.0 \times 8.9$ & 0.80$\sqrt{T/M}$ \\ \hline 
\end{tabular}
\end{center}
\caption{The history of cells used by HERMES.  The lengths of the
major and minor axes of the ellipse are indicated by $a$ and $b$; 
$T$ is the temperature of the cell in K, and $M$ is
the atomic mass of the gas in amu.}
\label{cellsize}
\end{table}
%%%%%%%%%%%%%%%%%%%%%%%%%%%%%%%%%%%%%%%%%%%%%%%%%%%%%%%%%%%%%%%

The storage cell is mounted onto cooling rails, which are 
cooled by temperature regulated gaseous helium. The optimum
operating temperature for hydrogen targets was found to be about 100\,K, where 
recombination and depolarization effects are low. Additionally, the lower
conductance at this temperature leads to a higher target thickness by a
factor $\sqrt{3}$ compared to room temperature. For deuterium 
gas the target 
temperature could be decreased further to about 65\,K, which increased the
target thickness by about 23\,\% compared to 100\,K.

The temperature of the cooling rails is monitored at three locations
by platinum resistors. Since 1999 an additional thermocouple wire has been 
attached to the aluminum foil close to the cell ellipse to directly monitor 
the cell temperature during injection of the HERA beam when a temperature 
rise of about 5--10\,K is usually detected.
The cooling rails
are mounted on the target chamber flange on the upstream end. 
Due to thermal contraction of the cell when being cooled, the position of the
cell center changes relative to the atomic beam source and the Breit-Rabi
polarimeter. From room temperature down to 100\,K the cell center moves by
0.9 mm. The alignment of the atomic beam source and Breit-Rabi polarimeter
relative to the cell center includes an offset to compensate for this
effect at the target operating temperature.

The cell is shown along with its support flange in Fig.~\ref{fig:cell}.
Two side tubes are connected to the beam tube, one to inject the polarized
atoms and another smaller 
one to sample about $5\%$ of the target gas
and analyze it in the target gas analyzer and the Breit-Rabi polarimeter.
The sample tube is tilted by $120^{\circ}$
with respect to the axis of the injection tube to
ensure that the sampled atoms have thermalized with the storage cell wall.
A permanent 
dipole magnet around the sampling tube outside the target chamber 
extends
the magnetic holding field of the target magnet to prevent depolarization
(see Fig.~\ref{fig:chamber}). 
A capillary near the location of the
feed tube allows injection of unpolarized gas into the cell.

\subsection{Atomic Beam Source}

\begin{figure*}
\begin{center}
\includegraphics[width=\textwidth]{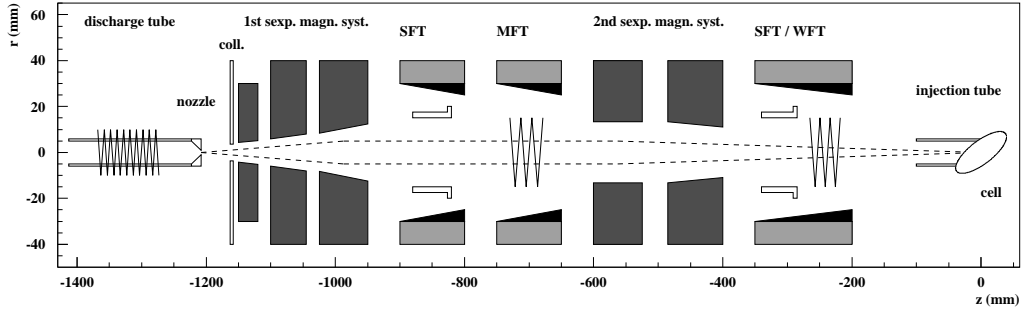} 
\end{center}
\caption{Schematic view (from downstream of the electron beam) of the
HERMES ABS with dissociator and collimator for beam formation. 
Two sets of sextupole magnets are located along the beam axis as are the
high-frequency transitions. The axis of the ABS is tilted by $30^\circ$ 
downwards with respect to the horizontal plane.}
\label{fig:abs}
\end{figure*}

The atomic beam source~\cite{Hermes:ABS} 
consists of a dissociator, a powerful differential pumping system, 
a beam forming
system, a sextupole magnet system to
focus atoms with $m_S$ = +$\frac{1}{2}$ into the storage cell,
and adiabatic high-frequency transitions to 
manipulate the hyperfine population of the atomic beam. 
Injected fluxes of
$\Phi^{ABS}$ $\approx$ $6.5 \times10^{16}$ atoms/s in the case of hydrogen 
(2 states) and
$\Phi^{ABS}$ $\approx$ $4.5 \times10^{16}$ atoms/s 
in the case of deuterium (3 states) have been
observed. These values are calculated from the density-dependent
spin exchange relaxation~\cite{Baumgarten:thesis2}. The injected nuclear
polarization $P^{inj}_z$ was constantly above 0.97 for hydrogen and above 0.91 
for deuterium. A schematic diagram of the ABS is depicted in 
Fig.~\ref{fig:abs}.

Molecular hydrogen/deuterium gas enters the dissociator from the left. 
The molecules are dissociated by a radio frequency discharge with a frequency
of 13.56 MHz in a pyrex tube, producing a degree of 
dissociation
up to 80\,\% ($\mathrm{H_2}$, $\mathrm{D_2}$) 
at a flow of about 1\,mbar\,l\,$\mathrm{s^{-1}}$ and a
radio frequency power of 300 W. To maximize the stability and the 
degree of dissociation, an amount of oxygen between 0.1 and 0.3 volume
percent is added to the molecular gas.

A high degree of dissociation at higher flow can potentially be achieved using
a recently developed microwave (MW) dissociator~\cite{koch},
which operates at a frequency of 2.45 GHz. With typical flows between 
1 and  2\,mbar\,l\,$\mathrm{s^{-1}}$ and a microwave power
of about 600\,W, a degree of dissociation higher than 80\,\% is achieved.
The MW dissociator was used in the year 2000 when 
running with deuterium, when it provided a target thickness improvement
of 5--10\,\%. 
The RF dissociator has been used since, because the ABS sextupole system 
is not optimized for the lower atomic velocity distribution 
produced by the MW dissociator, eliminating the possible benefit for hydrogen.

The atomic gas flows through a conical nozzle with an opening diameter of
2\,mm, which is cooled to about 100\,K. The water produced in the discharge 
freezes in the nozzle.  Although this beneficially reduces surface 
recombination, the ice layer continuously thickens and slowly reduces 
the gas conductance of
the nozzle. After typically 3--5 days of operation, the pressure in the 
dissociator tube rises substantially, increasing volume recombination
in the tube and resulting in a noticeable decrease in the degree 
of dissociation of gas
exiting the nozzle. Full recovery is achieved by slightly warming up the
nozzle to remove the ice layer.

After passing the cold nozzle, the gas expands into the vacuum of the 
dissociator chamber. A powerful differential pumping system with a total 
nominal pumping speed of more than 15000\,l\,$\mathrm{s^{-1}}$ 
suppresses scattering of the atomic beam, and ensures 
low gas flow into the HERA vacuum system. 

The magnet system of the HERMES atomic beam source consists of a total of
five Halbach type~\cite{halbach}  
segmented permanent sextupole magnets with a maximum poletip field of
about 1.5\,T. Every magnet consists of 24 segments 
of {\it{Vacodym}} alloy\footnote{Vacodym is a registered trademark of Vacuumschmelze GmbH, 
Hanau, Germany}. 
Four high-frequency transition units are available: two
between the sextupole magnets
and two after the last magnet. The system provides enough flexibility
that many different combinations of
 hyperfine states can be injected into the cell.
Between the sextupoles, a strong field transition (SFT) and a medium field
transition (MFT) can be used simultaneously, whereas the weak field
transition (WFT) and the SFT after the sextupole magnets can only be used 
separately, because they share the coils for the static and gradient magnetic
fields. The complete set of possible hyperfine populations that
can be produced is shown in Tab.~\ref{tab:injections}.
\begin{table*}[t]
  \begin{center}
      \begin{tabular}[b]{|c|c|c|c|c|c|c|c|}
        \hline
        \rule[-0.3cm]{0cm}{0.8cm}
        Gas & HFT (betw. 6-poles) & HFT (after 6-poles) & Inj. States & $P_e$ & $P_z$ & $P_{zz}$ & Use \\
        \hline\hline
        \rule[-0.2cm]{0cm}{0.6cm}
        & - & - & $\ket{1}$, $\ket{2}$ & +1 & 0 & - & Cal\\
        \rule[-0.2cm]{0cm}{0.6cm}
        {\bf\large H} & - & WFT 1-3 & $\ket{2}$, $\ket{3}$ & 0 & -1 & - & Data\\
        \rule[-0.2cm]{0cm}{0.6cm}
        & - &SFT 2-4 & $\ket{1}$, $\ket{4}$ & 0 & +1 & - & Data\\
        \rule[-0.2cm]{0cm}{0.6cm}
        & SFT 2-4 / MFT 2-3 & -  & $\ket{1}$ & +1 & +1 & - & Cal\\
        \rule[-0.2cm]{0cm}{0.6cm}
        & WFT 1-3 / MFT 1-3 & -  & $\ket{2}$ & +1 & -1 & - & Cal\\
        \rule[-0.2cm]{0cm}{0.6cm}
        & SFT 2-4 / MFT 2-3 & WFT 1-3  & $\ket{3}$ & -1 & -1 & - & Cal\\
        \rule[-0.2cm]{0cm}{0.6cm}
        & WFT 1-3 / MFT 1-3 & SFT 2-4  & $\ket{4}$ & -1 & +1 & - & Cal\\
        \rule[-0.2cm]{0cm}{0.6cm}
        & WFT 1-3, SFT 2-4 & -  & {\it{no state}} & - & - & - & Cal\\
        \hline
        \rule[-0.2cm]{0cm}{0.6cm}
        & - & - & $\ket{1}$ $\ket{2}$ $\ket{3}$ & +1 & 0 & 0 & Cal\\
        \rule[-0.2cm]{0cm}{0.6cm}
        {\bf\large D} & SFT 2-5 & WFT 1-4 & $\ket{3}$ $\ket{4}$ & 0 & -1 & +1 & Data\\
        \rule[-0.2cm]{0cm}{0.6cm}
        & SFT 3-5 & SFT 2-6 & $\ket{1}$ $\ket{6}$ & 0 & +1 & +1 & Data\\
        \rule[-0.2cm]{0cm}{0.6cm}
        & MFT 1-4 & SFT 3-5 & $\ket{2}$ $\ket{5}$ & 0 & 0 & -2 & Data\\
        \rule[-0.2cm]{0cm}{0.6cm}
        & MFT 1-4 & SFT 2-6 & $\ket{3}$ $\ket{6}$ & 0 & 0 & +1 & Data\\
        \rule[-0.2cm]{0cm}{0.6cm}
        & MFT 3-4, SFT 2-6 & - & $\ket{1}$ & +1 & +1 & +1 & Cal\\
        \rule[-0.2cm]{0cm}{0.6cm}
        & WFT 1-4, SFT 2-6 & - & $\ket{2}$ & +1 & 0 & -2 & Cal\\
        \rule[-0.2cm]{0cm}{0.6cm}
        & WFT 1-4, SFT 3-5 & - & $\ket{3}$ & +1 & -1 & +1 & Cal\\
        \rule[-0.2cm]{0cm}{0.6cm}
        & MFT 3-4, SFT 2-6 & WFT 1-4 & $\ket{4}$ & -1 & -1 & +1 & Cal\\
        \rule[-0.2cm]{0cm}{0.6cm}
        & WFT 1-4, SFT 3-5 & SFT 3-5 & $\ket{5}$ & +1 & 0 & -2 & Cal\\
        \rule[-0.2cm]{0cm}{0.6cm}
        & WFT 1-4, SFT 2-6 & SFT 2-6 & $\ket{6}$ & -1 & +1 & +1 & Cal\\
        \hline
      \end{tabular}
    \end{center}
\caption{Injection modes of the atomic beam source. The table shows the 
high frequency transition (HFT) units employed (2$^{nd}$ and 3$^{rd}$ columns), 
the hyperfine states injected in the target cell 
and the resulting electron ($P_e$), nuclear ($P_z$) and tensor ($P_{zz}$) 
polarizations in the ideal case of 100\,\% efficiency of the sextupole system 
and transition units, no depolarization inside the target cell
and infinitely strong guide field. The last 
column clarifies the purpose of each mode (Cal: calibration, monitoring; 
Data: normal HERMES operation). The injection modes for hydrogen 
and deuterium single states are listed separately. For some modes
 the MFT between the sextupoles has to be tuned as a WFT 1-4. The 
{\it{no state}} injection mode is possible only for hydrogen 
and has been operational since April 2002. The adopted labeling for the 
hyperfine states is explained in the Breit-Rabi diagram of 
Fig.~\protect\ref{breitrabi}.}
\label{tab:injections}
\end{table*}
\begin{figure*}
\begin{center}
\includegraphics[width=\textwidth]{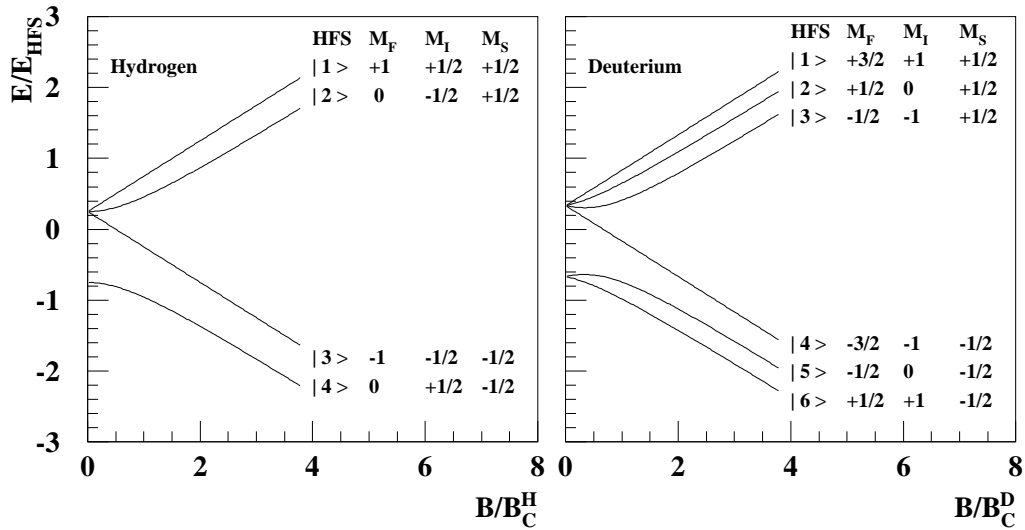} 
\end{center}
\caption{Hyperfine energy levels of hydrogen (left) and deuterium (right)
atoms as a function of the magnetic holding field (Breit-Rabi diagram) and
corresponding labeling. The field values are scaled with the corresponding 
critical field and energy values with the corresponding hyperfine energy.}
\label{breitrabi}
\end{figure*}

\subsection{Target Gas Analyzer}
\label{sec:TGA}

\begin{figure*}
\begin{center}
\includegraphics[width=\textwidth]{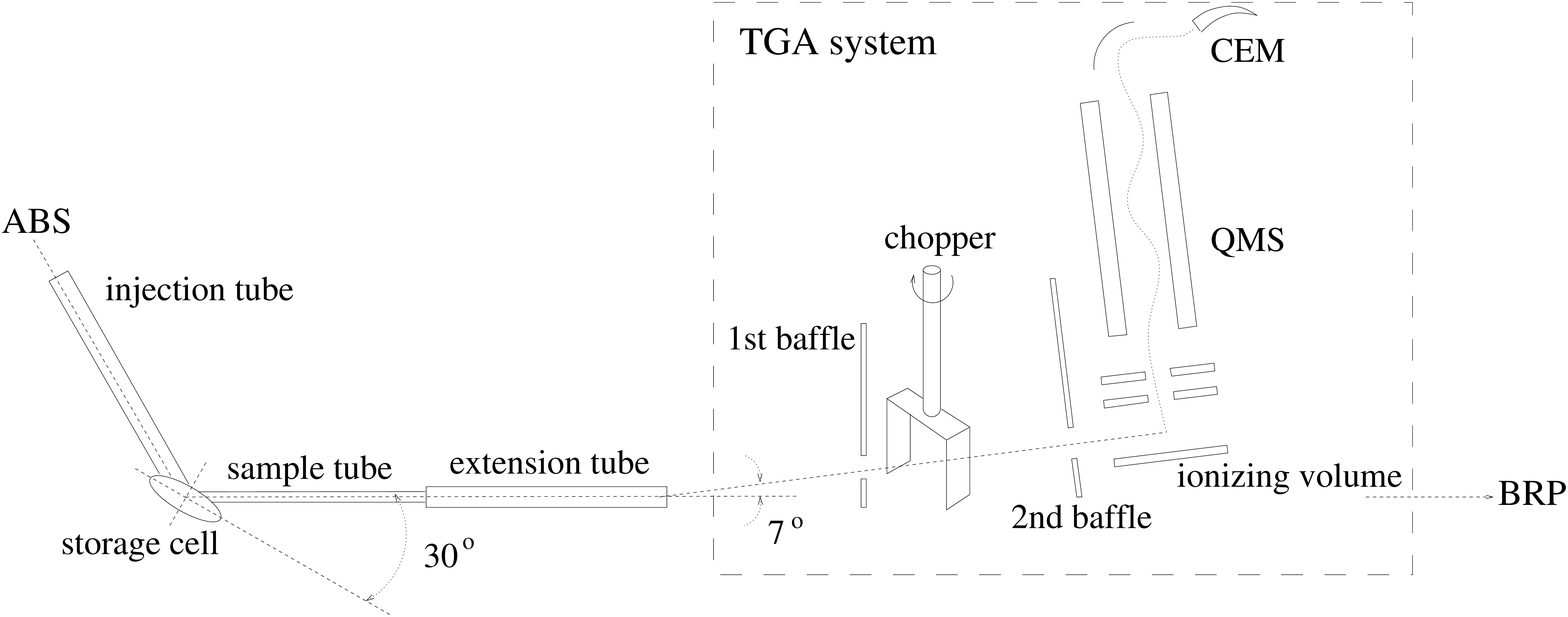} 
\end{center}
\caption{A schematic diagram of the Target Gas Analyzer. The gas of atoms 
and molecules diffuses through the extension tube and is collimated
by two baffles before entering the ionizing volume. 
The chopper is used for background subtraction}
\label{fig:tga}
\end{figure*}

\begin{figure*}
\begin{center}
\includegraphics[width=\textwidth]{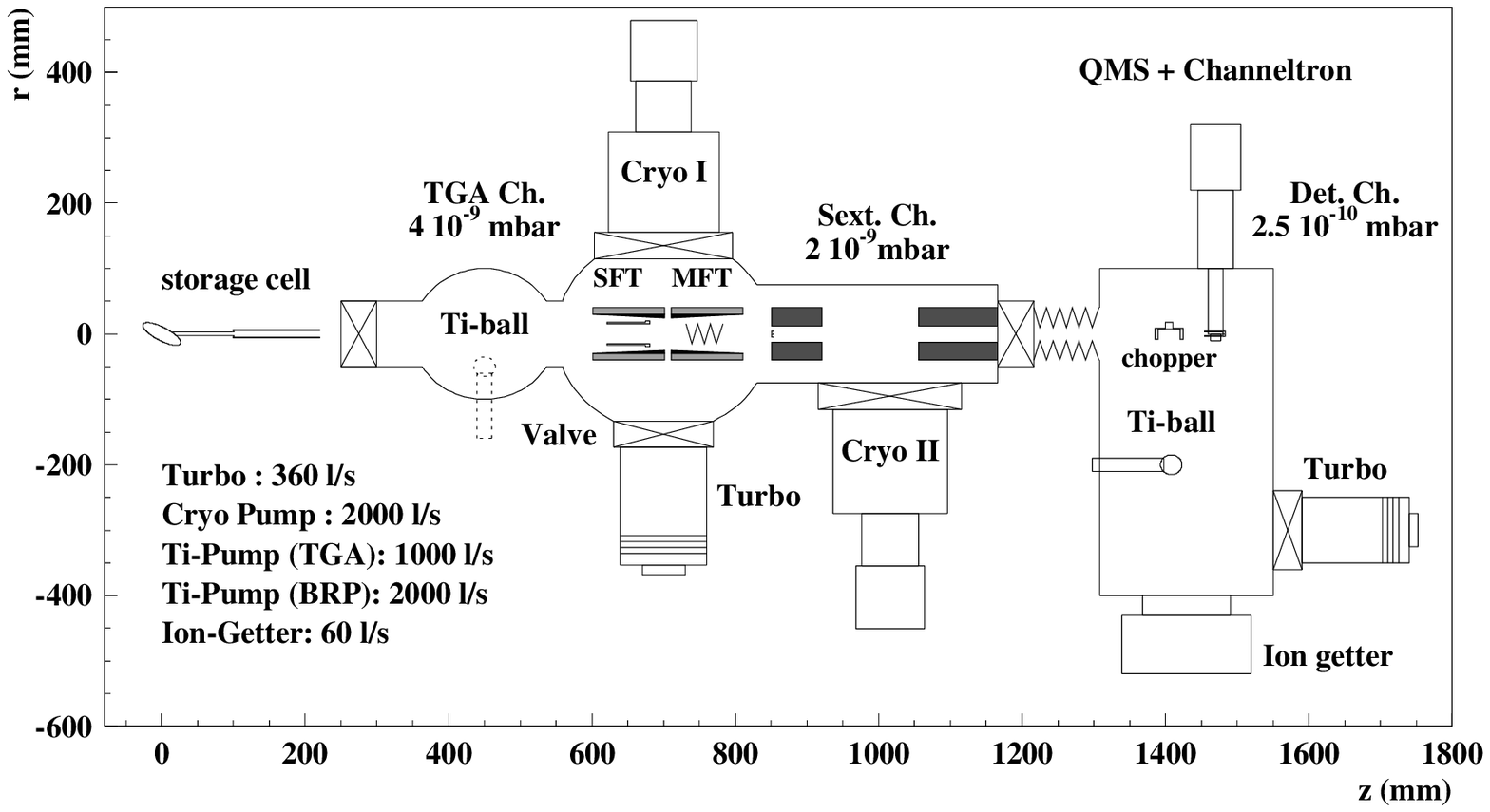} 
\end{center}
\caption{Schematic view of the BRP/TGA vacuum system.}
\label{fig:brp_vac}
\end{figure*}

The target gas analyzer (TGA)~\cite{Hermes:TGA} measures the atomic and 
molecular content of the gas extracted from the storage cell through
the sample tube.
The TGA arrangement, shown in Fig.~\ref{fig:tga}, consists of a pair of baffles,
 a chopper, a 90$^{\circ}$ off-axis quadrupole mass spectrometer (QMS) 
with a cross beam ionizer and a channel electron multiplier (CEM) 
for single ion detection. 
The TGA is integrated into the vacuum system of the 
BRP sextupole chamber, which is served by two cryopumps and a titanium 
sublimation pump, with a total pumping speed of about 7000\,$\mathrm{l\,s^{-1}}$. 
During operation, the pressure in the TGA detector is
about 4$\cdot$10$^{-9}\,$mbar. 
Prior to normal operation, the 
TGA vacuum chamber is baked with temperatures 
up to 180$\,^{\circ}$C for 48 hours. 
The BRP/TGA vacuum scheme is shown in Fig.~\ref{fig:brp_vac}. 

\begin{figure*}
\begin{center}
\includegraphics[width=\textwidth]{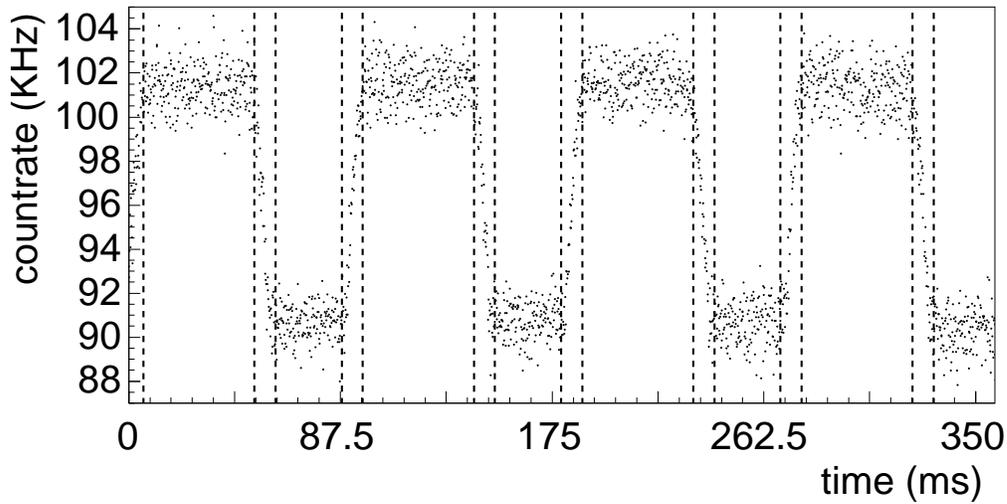} 
\end{center}
\caption{Time spectrum of atomic count rates in the TGA. 
The vertical lines indicate the binning used to define 
the chopper positions open, undefined and closed.}
\label{fig:tgaspec}
\end{figure*}

The TGA is mounted 7$^{\circ}$ off-axis with respect to the BRP, 
in order to not interfere with the beam entering the polarimeter. 
The first baffle ensures that only gas from the sample beam can reach 
the detector, while the second one acts as a collimator, so that 
no atoms can hit the metal parts of the ionizer and recombine. The chopper, 
placed between the two baffles, rotates at a frequency of 
5.5$\,$Hz, thus periodically 
blocking the sample beam in order to allow subtraction of the
 residual gas signal. 
Particles entering the detector are ionized by 70$\,$eV electrons, 
mass filtered with the QMS, and finally detected by the CEM. 
The pulses are counted by a 2048-bin
time-resolving counter (TRC). The bin length is typically set to 175$\,\mu$s
so that two chopper turns are collected within the whole bin range as shown 
in Fig.~\ref{fig:tgaspec}. The synchronization between chopper status 
and counting rate is given by a trigger signal coming from the chopper
motor. The beam rate is calculated by taking 
the difference between the number of counts collected with the 
chopper open and closed.

\subsection{Breit-Rabi Polarimeter}
\label{sec:brp}

The Breit-Rabi polarimeter (BRP)~\cite{Hermes:BRP} 
measures the relative populations
 $n_i$ of the hyperfine states of hydrogen (or deuterium) atoms contained 
in the sample beam. From this measurement, the absolute atomic polarizations 
can be calculated by applying the knowledge of the target magnetic field 
strength.

\begin{figure*}
\begin{center}
\includegraphics[width=\textwidth]{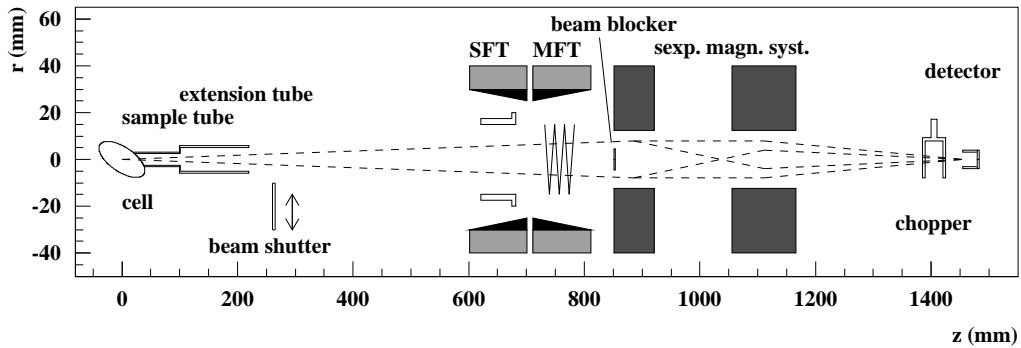} 
\end{center}
\caption{ Schematic layout of the BRP. The rf-transition units extend from
z = 600 to 810 mm. The beam blocker at the entrance to the first
6-pole magnet ensures 100\,\% rejection of atoms with  $m_s=-\frac{1}{2}$.
The beam shutter is used to measure the 
hydrogen contribution coming from dissociative water ionization.}
\label{fig:brp}
\end{figure*}

A schematic view of the BRP is shown in 
Fig.~\ref{fig:brp}. From left to right, the sample beam
leaves the extension tube of the target cell,
encountering first two hyperfine transition units, 
then a sextuple magnet system and eventually the detector stage. 
During operation, a differential pumping system keeps the pressure at 
2$\times$10$^{-9}\,$mbar in the sextupole chamber and at 
2.5$\times$10$^{-10}\,$mbar
in the detector chamber.

Two transition units are 
used to exchange the populations between pairs of hyperfine states:
a strong field transition unit (SFT) with 
tilted resonator that can be tuned for both $\pi$ and $\sigma$ transitions, 
and a medium field transition (MFT) unit that can induce various $\pi$ 
transitions according to the static field strength and gradient setting used.
%The sextupole system is composed of two magnets and spin-filters the sample 
%beam by focusing atoms with $m_s=+\frac{1}{2}$ towards the BRP geometrical 
%axis and defocusing atoms with $m_s=-\frac{1}{2}$.
The sextupole system is composed of two
magnets which focus the atoms of the sample beam with $m_s=+\frac{1}{2}$ 
towards the BRP geometrical axis and defocus the ones with $m_s=-\frac{1}{2}$.
A 9$\,$mm diameter beam blocker placed in front of 
the first sextupole magnet ensures that no atoms in $m_s = -\frac{1}{2}$ 
states can reach the detector.
The sextupole system
was redesigned in the year 2000 in order to improve the transmission 
probability at the typical cell temperature.
The detector stage is identical to the one employed for 
the target gas analyzer: a cross beam ionizer, a quadrupole mass 
spectrometer (QMS) and a channel electron multiplier (CEM). In contrast 
to the TGA, only hydrogen (or deuterium) {\it{atoms}} are detected by the BRP. 
The chamber residual gas correction is carried out by using a chopper in 
the same way as described for the TGA. An additional beam shutter is placed 
just after the extension tube exit to measure the possible 
contribution of atomic hydrogen coming from the dissociation of 
$\mathrm{H_2O}$ 
and $\mathrm{H_2}$ 
($\mathrm{D_2O}$, $\mathrm{DHO}$, $\mathrm{D_2}$ and $\mathrm{DH}$ 
in the deuterium case) present 
on the vacuum chamber walls\footnote{The beam shutter was installed at 
the beginning of the deuterium running period in 1998. During the 
1996/'97 hydrogen running period, the contribution of dissociative hydrogen
ionization was taken into account by detecting the 
intensity of masses 2 and 18\,amu 
in addition to mass 1\,amu~\cite{Baumgarten:thesis2}.}.

\section{Analysis of TGA and BRP signals}

\subsection{Interpretation of the TGA signals}
\label{sec:TGAsig}

The measured TGA atomic (molecular) signal $S_a$ ($S_m$) is proportional 
to the product of the {\it{particle}} 
flux $\Phi_a$ ($\Phi_m$) into the TGA, the mean inverse velocity 
of the particles $\bigl<\frac{1}{v_a}\bigr>$ ($\bigl<\frac{1}{v_m}\bigr>$),
the ionization 
cross section $\sigma_a^{ion}$ ($\sigma_m^{ion}$) and the detection 
probability $\epsilon_a$ ($\epsilon_m$). 
A dependence of the signal on the emission current is measured 
and taken into account by the normalization function $f(I_{emiss})$. 
Taking all parameters into account one finds:
\begin{equation}
S_{a,m}\propto f(I_{emiss})\,\,\sigma_{a,m}^{ion}\,\,\epsilon_{a,m}\,\,
\biggl<\frac{1}{v_{a,m}}\biggr>\,\,\Phi_{a,m}.
\label{eq:TGA1}
\end{equation}
The total flux $\Phi^{TGA}$ of H or D nucleons into the TGA
can be related to the particle fluxes $\Phi_a$ and 
$\Phi_m$ by:
\begin{equation}
\Phi^{TGA}=\Phi_a+2\Phi_m.
\label{eq:TGA2}
\end{equation}
The TGA geometrical acceptance is such that detected particles 
have bounced typically more than 100 times on the extension tube 
inner surface~\cite{Baumgarten:thesis2}. 
For this reason, particles entering 
the TGA detector have thermalized with the extension tube wall,
which is a few tens of degrees K warmer than the cell.
For thermalized atoms and molecules one has:
\begin{equation}
\biggl<\frac{1}{v_{m}}\biggr>=\sqrt{2}\,\biggl<\frac{1}{v_{a}}\biggr>.
\label{eq:TGA3}
\end{equation}
Using eqs. (\ref{eq:TGA1}), (\ref{eq:TGA2}) and (\ref{eq:TGA3}), it can be
shown that the partial beam rates $S_a$ and $S_m$ 
can be combined to form the variable $S_{tot}$ which is proportional
to $\Phi^{TGA}$:
\begin{equation}
S_{tot} = S_a + \kappa\sqrt{2} S_m \propto\Phi^{TGA},
\label{stot}
\end{equation}
where $\kappa$ is the calibration constant of the TGA giving 
the relative sensitivity of the TGA for atomic and molecular beams:
\begin{equation}
\kappa=\frac{\sigma_{a}^{ion}\,\,\epsilon_{a}}{\sigma_{m}^{ion}\,\,\epsilon_{m}}.
\label{eq:kappa}
\end{equation}

Eq.~(\ref{stot}) applies to the case of constant extension tube temperature
$T_{ext}$ and emission current $I_{emiss}$. Taking into account the 
dependence of $S_a$ and $S_m$ on the measured\footnote{A thermocouple on 
the extension tube was first 
installed at the beginning of 1999. In the 1997 analysis, the extension 
tube temperature was related to the cell temperature using the expression 
$T_{ext}(T)=T_0+\frac{300-T_0}{300}T$, where $T_0$ is a constant 
estimated from calibration studies, and 
T is the measured cell temperature in K.}
$T_{ext}$ and $I_{emiss}$, one
can define a temperature and emission normalized nucleon flow rate 
$\phi_{tot}$ into the TGA ionizer,
which is proportional to the injected ABS intensity $\Phi^{ABS}$
(neglecting the small contribution from the target chamber rest gas):
\begin{equation}
\phi_{tot}=\frac{S_{tot}\sqrt{T_{ext}}}{f(I_{emiss})} \propto\Phi^{ABS}
\label{eq:phitot}\,.
\end{equation}
The quantity $\phi_{tot}$ is a good monitor of ABS stability over time, 
as it is independent of recombination inside the storage cell. Following 
eq.~(\ref{eq:phitot}) one can also express
normalized nucleon flow rates for atoms, $\phi_a$, and for molecules, $\phi_m$,
by:
\begin{equation}
\phi_{a}=\frac{S_{a}\sqrt{T_{ext}}}{f(I_{emiss})}
\label{eq:phia}
\end{equation}
\begin{equation}
\phi_{m}=\frac{\sqrt{2}\kappa S_{m}\sqrt{T_{ext}}}{f(I_{emiss})}.
\label{eq:phim}
\end{equation}
The degree of dissociation of the sample beam, defined as the fraction 
of nucleons in atoms relative to all nucleons entering the TGA, 
is given by
\begin{equation}
\alpha^{TGA} \equiv \frac{\phi_a}{\phi_a+\phi_m} =
\frac{S_a}{S_a+\sqrt{2}\,\kappa\,\, S_m}.
\label{alpha_TGA}
\end{equation}
The degree of dissociation $\alpha^{TGA}$, also called atomic fraction, 
is displayed by the TGA roughly once per minute. 

\subsection{The calibration of the TGA}

The purpose of the TGA calibration is to determine the 
relative sensitivity $\kappa$ defined by eq.~(\ref{eq:kappa}). The calibration 
can be performed by varying the relative amount of atoms and molecules 
at a constant injected ABS total flux:
\begin{equation}
\phi_{tot}=\phi_a+\phi_m=const.
\label{eq:const_rate}
\end{equation}
Using the relations (\ref{eq:phia}) and (\ref{eq:phim}), one can write
\begin{equation}
\frac{\partial\Bigl(\frac{S_{m}\sqrt{T_{ext}}}{f(I_{emiss})}\Bigr)}
{\partial\Bigl(\frac{S_{a}\sqrt{T_{ext}}}{f(I_{emiss})}\Bigr)}=
-\frac{1}{\sqrt{2}\,\kappa}\,,
\end{equation}
where the minus sign reflects the fact that an increase in $\phi_m$ must
correspond to a decrease of $\phi_a$. Hence
a fit of the normalized molecular signal $\phi_m$ as function 
of the normalized atomic one $\phi_a$ 
results in a straight line with a slope of $-\frac{1}{\sqrt{2}\,\kappa}$.

Because 
the recombination process in the HERMES target is strongly temperature 
dependent, a controlled variation of the atomic fraction can be obtained 
by changing the temperature of the storage cell. 
A second possibility is to take advantage of an accidental beam loss 
in the region of the HERMES target which might temporarily change the 
surface property of the cell and cause a sharp drop of $\alpha^{TGA}$
followed by a slow recovery. 
The measured values for $\kappa$ are reported in Tab.~\ref{kappa_table}.
\begin{table}
\begin{center}
\begin{tabular}{|c|c|}
\hline
target/year & $\kappa$ \\ \hline \hline

$H_{||} (1997)$ & $0.551\pm0.026$ \\ \hline

$D_{||} (2000)$ & $0.861\pm0.024$ \\ \hline

$H_{\perp} (2002)$ & $0.55\pm0.05$ \\ \hline

\end{tabular}
\end{center}
\caption{Summary of the values of TGA calibration constant
$\kappa$ for the years 1997-2002.}
\label{kappa_table}
\end{table}

A detailed description of the TGA calibration and its performance can be
found in Ref.~\cite{Hermes:TGA}.

\subsection{Interpretation of the BRP signals}

The atomic signal detected by the BRP depends on the hyperfine population 
in the sample beam, the transmission 
probabilities of the sextupole system and on the setting of the transition
 units. For a stable storage cell temperature, the 
transmission of the 
magnetic system is fixed. The transition units can be 
switched on and off, changing the intensity of the beam 
reaching the detector.
The BRP signal can be described by the following expression:
\begin{equation}
S_i=c\sum_a\,\Biggl(\sum_b\sigma_b T_{ab}^{i}\,\Biggr)I_a\,,
\label{eq:brp_signal}
\end{equation}
where $I_a$ is the sample beam intensity (in atoms/s) 
of the atoms in hyperfine state $\ket{a}$, $T_{ab}^{i}$ describes the 
exchange of hyperfine states by the transition 
unit\footnote{If two units are working, the matrix $T^{i}$ has to be replaced
 by $T_{ab}^{ij}=\sum_c T_{ac}^i T_{cb}^j$} $i$, 
$\sigma_b$ is the sextupole transmission probability for state $\ket{b}$, 
and $c$ is a proportionality factor\footnote{The expression 
(\ref{eq:brp_signal}) simplifies into $\,S_{off}=c\sum_a\sigma_aI_a\,$ 
in the case that no transition unit is switched on.}.
The goal of these measurements is to determine the four (six) sample beam
intensities, $I_a$, for the hydrogen (deuterium) sample beam.
Summing eq.~(\ref{eq:brp_signal}) over index {\it{b}} one obtains:
\begin{equation}
S_i=c\sum_aM_{ia}I_a\,,
\label{eq:meas_matrix}
\end{equation}
where the {\it{measurement matrix}} $M_{ia}$ includes information 
concerning transmission probabilities and RF-transition efficiencies.

Assuming that $M_{ia}$ is known, 
one can solve the system of equations (\ref{eq:meas_matrix})
for $I_a$ if, during a given ABS injection status, the BRP transition units 
are operated in at least four (six) different modes in the case of hydrogen
 (deuterium).
The solution is given by
\begin{equation}
I_a=c^{-1}\sum_iR_{ai}S_i\,,
\label{eq:intensities}
\end{equation}
where\footnote{If the number of collected signals is 
larger than the number of unknowns, one has to use the 
{\it{pseudo-inverse}}
matrix $R=(M^TG_SM)^{-1}M^TG_S$, where $G_S$ is the inverse covariance 
matrix of the measured signals. This procedure is equivalent to a fit of 
the intensities $I_a$ to the signals $S_i$.} $R=M^{-1}$. 
The total sample beam intensity, 
$I_{tot}$, is obtained by the taking the sum over all hyperfine states:
\begin{equation}
I_{tot}=\sum_aI_a\,.
\label{eq:Itot}
\end{equation}
Finally, the relative hyperfine populations $n_a$ are given by 
normalizing the intensities $I_a$:
\begin{equation}
n_a=\frac{I_a}{I_{tot}}.
\label{eq:normalization}
\end{equation}
The polarizations of the atoms in the sample beam can now be calculated from 
the target magnetic field strength, as expressed 
by the following vectorial forms for hydrogen and deuterium, respectively:
\begin{eqnarray} \nonumber
(P_e,P_z)^T&=&M_P^H(n_1,n_2,n_3,n_4)^T\\
(P_e,P_z,P_{zz})^T&=&M_P^D(n_1,n_2,n_3,n_4,n_5,n_6)^T\,, \nonumber
\end{eqnarray}
where the matrices $M_P^H$ and $M_P^D$ contain the target 
holding field strength. e.g. for hydrogen:
\begin{equation}
{M^H_P} = 
\left( \begin{array}{cccc}
-1 &  a &  1 &  -a \\
1 &  -a &  -1 &  a \\
\end{array} \right)\,,
\end{equation}
where a = $\frac{k}{\sqrt{1+k^2}}$, $k = B/B_c$, $B_c$ = 50.7 mT.

The statistical uncertainties of the measured signals $S_i$ are represented 
by the diagonal covariance matrix $C_S$. Hence, the covariance matrix 
$C_I$ of the intensities $I_a$ is given by:
\begin{equation}
C_I=RC_SR^T=\bigl(M^TC_S^{-1}M\bigr)^{-1}\,.
\end{equation}
The sum over all the elements of $C_I$ gives the statistical 
uncertainty $\Delta I_{tot}$ of the total intensity $I_{tot}$:
\begin{equation}
(\Delta I_{tot})^2=\sum_{ab}(C_I)_{ab}\,.
\end{equation}
%=
The covariance matrix $C_n$ of the hyperfine populations $n_a$ is
\begin{equation}
(C_n)_{ab}=\sum_{cd}\Biggl(\frac{\partial n_a}{\partial I_c}\Biggr)(C_I)_{cd} 
\Biggl(\frac{\partial n_b}{\partial I_d}\Biggr)\,, 
\end{equation}
while the covariance matrix of the polarization vector can be calculated 
in the following way:
\begin{equation}
C_P=M_P^{H,D}C_n\,(M_P^{H,D})^{T}\,.
\end{equation}
From $C_P$ one can extract the statistical uncertainty of the polarization 
measurement.

During normal running conditions, the entire automated procedure
to measure and calculate the polarization vectors and their uncertainties
took roughly 
60$\,$s (90$\,$s) for hydrogen (deuterium), resulting in a statistical
 uncertainty of about $\pm$0.013 for hydrogen. For deuterium 
the statistical uncertainty was about $\pm$0.031 for positive 
and $\pm$0.022 for negative nuclear polarization. After an optimization 
of the BRP sextupole system, and an increase in the diameter of the
sample tube, the statistical uncertainty of a 60$\,$s measurement
for hydrogen was reduced to less than 0.5\,\%.

\subsection{The BRP calibration}

In the previous section it was shown that the atomic polarizations can 
be calculated from the measured BRP signals assuming the knowledge of the 
measurement matrix $M_{ia}$. This includes the efficiencies of the transition
 units and the transmission probabilities of the sextupole system for 
various hyperfine states. The procedure used to determine these parameters
 is called the BRP calibration.

The basic idea of the calibration measurement is to make the system of 
equations (\ref{eq:meas_matrix}) overdetermined so that additional 
parameters like efficiencies or sextuple transmissions can be extracted
 through a fit to the data. 
Due to the fact that the transition units can be combined in
several possible ways, 
the number of signals which can be collected during a 
polarization measurement exceeds the number of hyperfine states $N_{hfs}$.
However, the number of unknowns that can be determined in this way is still
 less than the total number of parameters. The problem can be 
overcome by cycling this measurement through several
(total number $N_{ABS}$) injection modes (see Tab.~\ref{tab:injections})
so that an adequate number of signals can be collected.

\subsubsection{The transition efficiencies}

We first discuss the method to determine the transition efficiencies,
assuming {\em a priori} knowledge of the sextupole transmissions 
$\sigma_n$ for each hyperfine state.  Calibration of those transmissions 
will then be discussed in Sec.~\ref{sec:6ptr}.

The BRP contains two transition units, a strong field (SFT) and a medium
 field (MFT) unit. The SFT can drive two transitions for hydrogen 
($\ket{1}\leftrightarrow\ket{4}$ and $\ket{2}\leftrightarrow\ket{4}$) and 
five for deuterium. The efficiencies of the SFT transitions in the case of 
hydrogen are called $\varepsilon_{s14}$ and $\varepsilon_{s24}$.
The MFT is a multi-step transition unit which can drive two sub-transitions
 for hydrogen ($\ket{1}\leftrightarrow\ket{2}$ and 
$\ket{2}\leftrightarrow\ket{3}$\footnote{The order of the sub-transitions
 depends on the sign of the gradient field, which is negative in the 
HERMES BRP~\cite{Baumgarten:thesis2}.}) and three for deuterium. 
For hydrogen the MFT is tuned either as a $\ket{1}\leftrightarrow\ket{3}$
(which is achieved by applying the $\ket{1}\leftrightarrow\ket{2}$ and 
$\ket{2}\leftrightarrow\ket{3}$ transitions sequentially) or as a 
$\ket{2}\leftrightarrow\ket{3}$ only. In the latter case the MFT unit
is tuned to avoid the $\ket{1}\leftrightarrow\ket{2}$ sub-transition.
Although nearly zero, the residual $\ket{1}\leftrightarrow\ket{2}$ efficiency
for the $\ket{2}\leftrightarrow\ket{3}$ tune of the MFT has to be taken into
account.
\begin{table*}
\begin{center}
\begin{tabular}[b]{|c|c|c|c|c|c|}
        \hline
        \rule[-0.3cm]{0cm}{0.8cm}
        \hspace{4mm} BRP-Signal \hspace{4mm} & \hspace{2mm} Name \hspace{2mm} &\hspace{2mm} MFT-RF \hspace{2mm} & \hspace{2mm} MFT \hspace{2mm} & \hspace{2mm} SFT-RF \hspace{2mm} & \hspace{2mm} SFT \hspace{2mm}\\
        \hline\hline
        \rule[-0.2cm]{0cm}{0.6cm}
        OFF & off & OFF & OFF & OFF & OFF \\
        \hline
        \rule[-0.2cm]{0cm}{0.6cm}
        SFT 1-4 & s14 & OFF & OFF & ON & 1-4 \\
        \rule[-0.2cm]{0cm}{0.6cm}
        SFT 2-4 & s24 & OFF & OFF & ON & 2-4 \\
        \hline
        \rule[-0.2cm]{0cm}{0.6cm}
        MFT 1-3 & m13c14 & ON & 1-3 & OFF & 1-4 \\
        \rule[-0.2cm]{0cm}{0.6cm}
        MFT 1-3 & m13c24 & ON & 1-3 & OFF & 2-4 \\
        \rule[-0.2cm]{0cm}{0.6cm}
        MFT 2-3 & m23c14 & ON & 2-3 & OFF & 1-4 \\
        \rule[-0.2cm]{0cm}{0.6cm}
        MFT 2-3 & m23c24 & ON & 2-3 & OFF & 2-4 \\
        \hline
        \rule[-0.2cm]{0cm}{0.6cm}
        MFT 1-3/SFT 1-4 & m13s14 & ON & 1-3 & ON & 1-4 \\
        \rule[-0.2cm]{0cm}{0.6cm}
        MFT 1-3/SFT 2-4 & m13s24 & ON & 1-3 & ON & 2-4 \\
        \rule[-0.2cm]{0cm}{0.6cm}
        MFT 2-3/SFT 1-4 & m23s14 & ON & 2-3 & ON & 1-4 \\
        \rule[-0.2cm]{0cm}{0.6cm}
        MFT 2-3/SFT 2-4 & m23s24 & ON & 2-3 & ON & 2-4 \\
        \hline
      \end{tabular}
     \end{center}
    \caption{Possible switching modes of the BRP transition units 
for hydrogen running. During a calibration 
measurement all signals are collected, while for a polarization measurement
 only a subset of them is chosen.}
\label{tab:brp_signals}
\end{table*}

Due to the limited space available for the BRP equipment, there is 
cross-talk between the two BRP transition 
units when both static fields are switched on, so that the 
tuning of the MFT is influenced by the SFT field,
whereas the SFT is not influenced by the (low) MFT field. The MFT
 efficiencies have to be measured separately for each SFT setting, 
as they depend on the exact static field. For hydrogen this leads to an 
increase in the number of MFT efficiencies from four to eight. The various 
MFT efficiencies are identified by means of the subscripts $c14$ (for SFT 1-4
 static field on) or $c24$ (SFT 2-4 static field on) and by the 
subscripts r13 or r23 which indicate the particular sub-transition being
considered. For example, 
the notation $\varepsilon_{m13r23c24}$ represents the efficiency of the 
transition MFT 1-3, step 2-3, tuned together with the static field of 
the SFT 2-4.
The total number of efficiencies $N_{eff}$ is 10 for hydrogen and 41 
in the case of deuterium. The number $N_{BRP}$ of possible BRP signals related 
to various transition states is 11 for hydrogen (as shown 
in Tab.~\ref{tab:brp_signals}) and 29 for deuterium.

If the ABS is operated in $N_{ABS}$ different injection modes, and the 
BRP transition units switch through all possible $N_{BRP}$ states, the 
total number of collected signals is $N_s=N_{ABS}\, N_{BRP}$. In this way, 
recalling eq.~(\ref{eq:meas_matrix}), one obtains a system of $N_s$ equations
 containing $N_u=N_{eff}+N_{ABS}\, N_{hfs}$ 
unknowns\footnote{During the evaluation of the transition efficiencies, the 
sextuple system transmissions probabilities are assumed to be constant.}, 
where $N_{hfs}=4\,(6)$ in the case of hydrogen (deuterium). The system 
can be solved if $N_f=N_s-N_u\ge 0$, where $N_f$ 
counts the degrees of freedom and is given by
\begin{eqnarray} \nonumber
N_f&=&N_s-N_u \\
&=&(N_{BRP}-N_{hfs})\, N_{ABS}-N_{eff}\,.
\end{eqnarray}
In such a case, both the hyperfine state intensities $I_a$ and the 
transition efficiencies $\varepsilon$ can be determined. For hydrogen, 
the calibration measurement makes use of 7 ABS injection modes, 
resulting in $N_f=39$. In the case of deuterium two possible options have 
been developed, one with 5 ABS modes ($N_f=74$) and the other with 
6 ($N_f=97$).

The solution of the system is carried out by applying a Runge-Kutta 
method as explained in~\cite{Hermes:BRP}. The resulting covariance 
matrix $C_x$ has the form
\begin{equation}
C_x=\left( \begin{array}{cc}
              C_{\varepsilon}&C_{\varepsilon,I}  \\
              C_{I,\varepsilon}&C_{I} \\
  \end{array} \right)\,,
\end{equation}
where $C_{\varepsilon}$ contains uncertainties and covariances of the 
efficiencies and is used to calculate the systematic uncertainty of the 
polarization measurement. The resulting systematic uncertainty induced by
the determination of the transition efficiencies of the order of 1.5$\,\%$.
Tab.~\ref{tab:brp_eff} summarizes the transition efficiency measurements 
performed in 1997 with hydrogen. The efficiencies of the transitions
for deuterium running can be found in Ref.~\cite{Hermes:BRP}.
\begin{table*}
  \begin{center}
      \begin{tabular}[b]{|c|c|c|c|c|}
        \hline
        \rule[-0.3cm]{0cm}{0.8cm}
        Name & MFT & SFT & Efficiency (\%)& uncertainty (\%)\\
        \hline\hline
        \rule[-0.2cm]{0cm}{0.6cm}
        $\varepsilon_{s14}$ & - & 1-4 & 99.3 & 1.34\\ 
        \rule[-0.2cm]{0cm}{0.6cm}
        $\varepsilon_{s24}$ & - & 2-4 & 101.0 & 1.03\\        
        \hline
        \rule[-0.2cm]{0cm}{0.6cm}
        $\varepsilon_{m13r12c14}$ & 1-3 & 1-4 & 101.0 & 1.03\\
        \rule[-0.2cm]{0cm}{0.6cm}
        $\varepsilon_{m13r23c14}$ & & & 98.6 & 1.41\\
        \hline
        \rule[-0.2cm]{0cm}{0.6cm}
        $\varepsilon_{m23r12c14}$ & 2-3 & 1-4 & -1.0 & 1.13\\
        \rule[-0.2cm]{0cm}{0.6cm}
        $\varepsilon_{m23r23c14}$ & & & 94.9 & 1.23\\ 
        \hline
        \rule[-0.2cm]{0cm}{0.6cm}
        $\varepsilon_{m13r12c24}$ & 1-3 & 2-4 & 99.9 & 0.42\\
        \rule[-0.2cm]{0cm}{0.6cm}
        $\varepsilon_{m13r23c24}$ & & & 90.3 & 0.82\\
        \hline
        \rule[-0.2cm]{0cm}{0.6cm}
        $\varepsilon_{m23r12c24}$ & 2-3 & 2-4 & 0.0 & 1.10\\
        \rule[-0.2cm]{0cm}{0.6cm}
        $\varepsilon_{m23r23c24}$ & & & 97.4 & 1.04\\ 
        \hline
      \end{tabular}
   \end{center}
\caption{BRP transition unit efficiencies measured during the 1997 
hydrogen running fixing the sextupole transmission ratio to 
$r_{21}=1.029$. Each efficiency is shown along with 
an explanation of the MFT and SFT static field setting.}
    \label{tab:brp_eff}
\end{table*}

\subsubsection{The sextupole transmissions}
\label{sec:6ptr}

In the previous section it was shown how the efficiencies 
$\varepsilon_i$ of the BRP transition can be measured, assuming  
{\em a priori} knowledge of the sextupole transmissions $\sigma_n$ for each
 hyperfine state. The attempt to fit both $\varepsilon_i$ and all
of the $\sigma_n$ 
within the same algorithm fails. Nevertheless, the transmission probabilities
 can be determined by separating the fitting procedure into two steps, 
as shown below.

In the BRP (Fig.~\ref{fig:brp}), 
atoms with $m_S=-\frac{1}{2}$ are either defocused 
or stopped by the beam blocker, having therefore zero transmission 
probability ($\sigma_3=\sigma_4=0$ for hydrogen and 
$\sigma_4=\sigma_5=\sigma_6=0$ for deuterium). 
Moreover, due to the normalization (eq.~(\ref{eq:normalization})), 
only ratios of transmission probabilities are required for measuring the 
hyperfine populations. Thus the remaining required parameters are 
$\sigma_2/\sigma_1$ for hydrogen and $\sigma_2/\sigma_1$ and 
$\sigma_3/\sigma_1$ for deuterium. 
%Due to the smaller electron-nucleon 
%coupling ($B_C^D=$11.7$\,$mT against $B_C^H=$50.7$\,$mT), in the deuterium 
%case the Stern-Gerlach force depends entirely, in high fields, on $m_S$, 
%so that the ratios $\sigma_2/\sigma_1$ and $\sigma_3/\sigma_1$ for deuterium 
%can assumed to be one. 

The Stern-Gerlach force of the sextupole magnets is weaker for hydrogen atoms
in state $\ket{2}$ compared to state $\ket{1}$ for magnetic fields $B$ 
of the order or less than
the hydrogen critical field $B_c = 50.7$ mT,
near the axis of the sextupole system. Thus near-axis atoms in
hyperfine states $\ket{1}$ and $\ket{2}$ will experience a different
transmission efficiency and the transmission ratio 
$r_{12}=\sigma_{1}/\sigma_{2}$
is expected to exceed unity by a small amount. Due to the smaller
hyperfine coupling of deuterium, $B_c=11.7$ mT, the deviation
from unity of the ratio $\sigma_2/\sigma_1$ and $\sigma_3/\sigma_4$, is
expected to be significantly smaller than for hydrogen. Ray tracing 
simulations~\cite{Hermes:ABS}
show that the deviation is in fact negligible.
From these consideration it follows that the only 
transmission ratio that needs to be measured is $r_{21}=\sigma_2/\sigma_1$ 
for hydrogen.

From the BRP calibration the transition efficiencies are extracted as 
a function of the parameter $r_{21}$. The transmission ratio can thus be 
evaluated by minimizing the $\chi^2$ of the calibration results. As the 
transmission probability of the sextupole system depends on the velocity 
distribution of the atomic sample, the storage cell temperature $T_{cell}$ 
has to stay constant over the calibration measurement. This procedure applied 
to a calibration measurement taken at $T_{cell}=$95$\,$K has produced the 
following result~\cite{Hermes:BRP}:
\begin{equation}
\frac{\sigma_2}{\sigma_1}\Big{|}_{95\,K}=1.029\pm 0.0015\,.
\end{equation}
The small uncertainty of the measured ratio can be neglected in the calculation 
of the systematic uncertainty of the polarization measurement.
The potential cell temperature dependence of the transition efficiencies can 
be neglected as the velocity distribution of the sample is fixed by the 
BRP sextupole system. 
Therefore, a polarization measurement performed collecting 
at least one signal more than the minimum required number 
(as is always the case) can be used to fit the ratio $r_{21}$ for 
various cell temperatures.  

A more detailed description of the BRP calibration and performance can be found
in Ref.~\cite{Hermes:BRP}.

%%%%%%%%%%%%%%%%%%%%%%%%%%%%%%%%%%%%%%%%%%%%%%%%%%%%%%%%%%%%%%%%%%%%%%
%\input{interactions_bis.tex}

\section{Processes occuring inside the Storage Cell}

Inside the storage cell, the atoms of the target may experience several
interactions that have an influence on the average target polarization $P_T$.
Their complete understanding  is therefore fundamental for the reduction
 of the systematic uncertainty of the target polarization. 
In this section an overview of these
processes will be given, while a quantitative estimate of their
contributions for the various running periods of the HERMES target 
can be found in Sec.~\ref{sec:performance}.
The processes are divided into two categories --- recombination processes
and spin relaxation processes.

Recombination and relaxation by wall collisions are surface effects whereas
spin exchange collisions and resonant interaction of beam and target gas 
happen in the gas storage cell volume. The operating conditions 
of the target are chosen such that these unwanted effects are suppressed as
much as possible.
It should be noted that spin relaxation can be studied and monitored with good
statistics by means of the BRP, whereas the residual polarization after 
recombination can be studied only using a scattering process like deeply 
inelastic scattering with a much lower rate.

\subsection{Recombination Processes}

The atoms that are injected into the HERMES storage cell may recombine
to molecules, possibly keeping part of their nuclear polarization.
Recombination~\cite{Hermes:recomb}
can in principle take place either
on the storage cell walls (surface recombination) or in the gas phase
(volume recombination). At the low gas densities of the HERMES target,
volume recombination is negligible. On the cell walls two types of 
recombination processes with surface catalysis can be distinguished, which
are expected to result in different residual nuclear polarization
in the molecules:
\begin{itemize}
\item{the Langmuir-Hinselwood (L-H) process, where two adsorbed
thermalized atoms react with each other;}
\item{the Eley-Rideal (E-R) process, where an atom from the gas phase
reacts directly with an adsorbed atom~\cite{Hol94}.}
\end{itemize}
The residual polarization of the surface atoms recombining through the E-R 
mechanism in the cell was measured with the HERMES spectrometer, making 
use of a well known DIS asymmetry. The result was presented
in Ref.~\cite{Hermes:beta} and will be discussed in section 5.2.

%In Ref.~\cite{Hermes:recomb} by studying their temperature and 
%density dependence, three different recombination mechanisms have 
%been identified in HERMES cell: 
%(i) An activated E-R process in the case for temperatures
%above 120\,K. For lower temperatures L-H processes (ii) between two
%physisorbed atoms and (iii) between one physisorbed and one chemisorbed
%atom from the methyl group could be distinguished. 
%The working temperature of the cell at 100\,K for the case of hydrogen has 
%been set to minimize the recombination effects 
%(i) and (ii) + (iii) which present opposite temperature dependences.
%For the case of the running with deuterium in 2000, as it will be shown in 
%section 6.2, no recombination
%was detected and the working temperature could be decreased to 60\,K.
Since the storage cell temperature is usually kept below 140\,K, a small 
amount of water originating from the ABS dissociator 
freezes on the surface and grows into an ice layer on top
of the Drifilm surface. The presence of water is important since
water is known to inhibit hydrogen or deuterium recombination.
% The most likely reason for this is that the
% chemical binding energy of a hydrogen atom in water (H-OH) of
% 498 kJ$\mathrm{mol^{-1}}$ is higher than in a hydrogen molecule 
% (435.99 kJ/mol).
% The same applies for the case of deuterium, where the bond strength of
% D-OD is higher than that of D-D bonds.
Fresh storage cell coatings are hydrophobic, but
secondary particles from the HERA beam quickly change 
the chemical structure of the Drifilm surface, 
causing the surface to become less hydrophobic.
This was predicted to occur in the longitudinal holding field
by simulations of acceleration 
by the beam bunch electric fields of positive ions produced by 
the beam in the target gas.  Even when not at cyclotron resonance,
the ions stochastically gain energy
until they hit the cell wall with energies of order 100\,eV.
Surface doses in the range 10\,Gy/s can occur, depending on
the magnetic holding field and the beam current.
Furthermore, the downstream portion of the small cell used
since December 1999 is exposed to synchrotron radiation
that is single-scattered from collimator C1.
Fortunately, it has been observed that during normal operation, 
the quality of the cell surface gradually recovers and 
the amount of recombination in the cell decreases, thanks 
to the steady accumulation of ice on the surface.

A detailed description of the recombination studies performed on the HERMES
storage cell can be found in Ref.~\cite{Hermes:recomb}.

\subsection{Spin relaxation}
Three different atomic spin relaxation mechanisms can in principle be
 observed to occur in the HERMES
target, as they cause a change in the hyperfine population 
inside the storage cell.
One relaxation process takes place on the wall of the storage cell. 
Another is due to two-body spin-exchange collisions of atoms in the gas 
phase. In the third mechanism 
the high frequency fields generated by the passage of the very short
HERA beam bunches
can cause a depolarization in the target atoms under certain 
conditions~\cite{Hermes:beamdepol}. Spin relaxation of atomic hydrogen 
by wall or spin
exchange collisions with various types of wall coatings has been under
study for many years 
in the context of a hydrogen maser for low holding fields~\cite{steffens}. 
For the HERMES target further studies 
of the magnetic field dependence of spin
relaxation as well as studies of the temperature dependence and 
density dependence of the transition spectra have been carried 
out~\cite{Baumgarten:thesis2,bra95,pri93,pri94,kol98}.
The results of these studies are summarized in the following sections.

\subsubsection{Wall relaxation}
The atoms injected
by the atomic beam source experience several hundred
wall collisions during their diffusion process through the storage cell.
An attractive Van-der-Waals potential between the storage cell wall and a
diffusing atom causes the atom to be physisorbed, i.e. to stay for a certain
time close to the surface until it is desorbed again. The mean residence time
on the wall $\tau_s$ is described by the Arrhenius equation:
\begin{equation}
\tau_s=\tau_0\,e^{\bigl(\frac{E_b}{kT}\bigr)}\,,
\label{eq:arrenius}
\end{equation}
where $\tau_0$ is the high temperature limit of $\tau_s$, $E_b$ and $T$ 
are the adsorption energy and the temperature respectively, and $k$ 
is the Boltzmann constant.
During the time an adsorbed atom stays on the wall 
($\approx 10^{-10}$ s for the HERMES cell), the magnetic moment of 
its unpaired electron interacts with magnetic moments on the surface by means
of a dipolar coupling and/or with other unpaired electrons through exchange 
interactions~\cite{bou63a,bou63b,bou65,and63}.
Depending on the surface type, temperature and
magnetic field, one can define a probability that the atom leaves
the wall in a certain hyperfine state. This is described by means of
transition probabilities $W_{jk}$, each of which represents the probability to
find an atom in state $\ket{k}$ that was in state $\ket{j}$ prior to the wall collision.
The theoretical description and the calculation of the transition probability 
matrix for the HERMES target can be found in~\cite{Baumgarten:thesis2}.

\subsubsection{Spin Exchange Depolarization}

Atoms colliding with each other in the gas phase can undergo spin exchange,
where hyperfine transitions can occur via interaction between the
electron spins~\cite{pur56}. Due to the 
conservation of angular momentum, the sum of the magnetic quantum 
numbers $m_F$ before and after the spin exchange collision has to be the same:
\begin{equation}
\Delta\sum_im_F=0\,.
\end{equation}
One can see that if an initial population of the gas consists only
of pure state $\ket{1}$ or $\ket{3}$ in the case of hydrogen and $\ket{1}$
or $\ket{4}$ in the case of deuterium, no relaxation by means of spin exchange 
is possible.
This is actually not the case for the HERMES cell, where
to double the density, more than one atomic state 
is injected during normal operation.
The effect of spin exchange collisions on the hyperfine populations can be
described by a tensor $M_{jk}^i$, which gives the probability to find an
atom in state $\ket{i}$ after a collision of two atoms in the state
$\ket{j}$ and $\ket{k}$ prior to the collision:
% \begin{equation}
% \frac{dN_i}{dt} \propto \sum_{j,k} M_{jk}^i\,.
% \label{eq:sec}
% \end{equation}
The number of spin exchange collisions per unit time and volume is
\begin{equation}
\stackrel{\cdot}N_{se}=n^2 \sigma_{se} \erw{v_r}\,
\label{eq:secol}
\end{equation}
where $n$ is the gas density, $\sigma_{se}$ is the spin exchange cross 
section and $v_r$ is the relative velocity of the atoms. 
The density dependence of spin-exchange collision relaxation, combined with
the knowledge of $\sigma_{se}$~\cite{des76,all72}, 
was used to determine the density inside the HERMES 
cell~\cite{Baumgarten:thesis2}.  The results will be 
reported in Sec.~\ref{sec:performance}.

\subsubsection{Evolution equation for the description of the spin relaxation 
process}\label{sec:maseq}

\begin{figure*}
\begin{center}
\includegraphics[width=\textwidth]{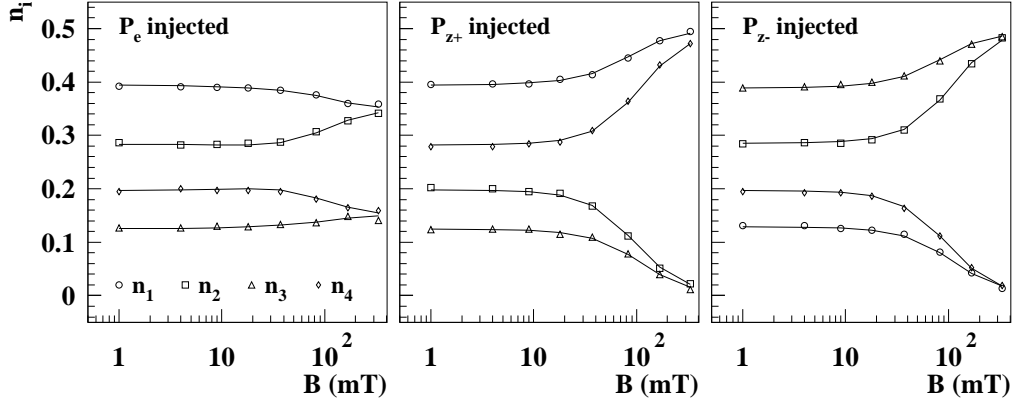} 
\end{center}
\caption{Measured fractional hyperfine populations vs. magnetic field for hydrogen 
for the various injection modes. The measured values are given by 
symbols, while the lines show a fit using the solutions of 
eq.~(\protect\ref{eq:master})}
{\label{fig:mfit}}
\end{figure*}

The various spin relaxations due to wall collisions and spin exchange collisions
cannot be treated separately. After an interaction with the surface during
a wall collision, an atom in the gaseous phase can then interact through spin
exchange collision and, at the next wall collision, interact again with the
surface. Since the atoms are constantly exposed to both interactions,
an appropriate description is given by an evolution equation
for the hypefine fractional populations
$n_i$ of the hyperfine state $\ket{i}$~\cite{Baumgarten:thesis2,bra95}\footnote
{In these references, this equation is designated as the {\it{master equation}}}:
% \begin{equation}
%\begin{multline}
%\frac{dN_i}{dt} = N_i^{inj} - \left\langle\frac{1}{\tau_d}\right\rangle N_i 
%   + \left\langle\frac{1}{\tau_f}\right\rangle \sum_k W_{jk}N_k \\  
%   + \left\langle\frac{1}{\tau_{se}}\right\rangle \sum_{jk} M_{jk}^i N_j N_k\,%.
%\label{eq:master}
%\end{multline}
% \end{equation}
% \begin{equation}
\begin{multline}
% \frac{dN_i}{dt} = \frac{1}{\tau_d} \left((N_i^{inj} - N_i) 
%    + \frac{\tau_d}{\tau_f}\sum_k W_{jk}N_k
%    + \frac{\tau_d}{\tau_{se}}\sum_{jk} M_{jk}^i N_j N_k\right).
\frac{\mathrm{d}n_i}{\mathrm{d}t} = \frac{1}{\tau_d} (n_i^{inj} - n_i) 
   + \frac{1}{\tau_f}\sum_k R_{jk}n_k 
   + \frac{1}{\tau_{se}}\sum_{jk} M_{jk}^i n_j n_k.
\label{eq:master}
\end{multline}
% \end{equation}
The first two terms of eq.~(\ref{eq:master}) characterize the incoming 
and outgoing fluxes, the third term represents the wall relaxation and the last term
the spin exchange relaxation.
The effect of recombination is not included, and it is assumed that the total
flux from the ABS is constant with time. 
In eq.~(\ref{eq:master}), three time constants appear: 
$\tau_d$ (the diffusion time) is the mean occupation time of atoms within
the storage cell, $\tau_f$ is the mean time of flight between two
wall collisions and $\tau_{se}$ is the mean time of flight between two spin
exchange collisions. 
For times that are large compared to $\tau_d$, the relative hyperfine populations 
converge toward a dynamic equilibrium ({\it{steady state}}).
% The steady state population is of practical interest, because the BRP
% requires a stable hyperfine population for a polarization measurement. 
%To properly describe the spin-relaxation inside the cell, 
%the cell surface has been subdivided into slices
%and eq. \ref{eq:master} applied to each slice. Proper boundary 
%conditions between neighbouring slices have been applied to satisfy 
%continuity.
As an example of the application of eq.~(\ref{eq:master}), 
the solution of the system of linear equations in the steady state
(with $\frac{\mathrm{d}n_i}{\mathrm{d}t} = 0$)
was solved directly in the case of hydrogen or with the help
of a numerical iteration algorithm in the case of 
deuterium~\cite{Baumgarten:thesis2}.
Fig.~\ref{fig:mfit} represents the measured hyperfine fractional 
populations as a function of the magnetic field for hydrogen and a
fit to the data using the solutions of eq.~(\ref{eq:master}).
The fitted parameters are the three time constants and optionally
the injected atomic state fractional populations.  The $R$ and $M$ 
transition probability matrices
are calculated on the basis of theoretical models~\cite{Baumgarten:thesis2}.
These data show that a holding field greater than 300\,mT is needed 
to have sizable polarization for the case of two injected states.

\subsubsection{Bunch field induced depolarization}
\label{sec:bunch_depol}

\begin{figure*}
\begin{center}
\includegraphics[width=0.85\textwidth]{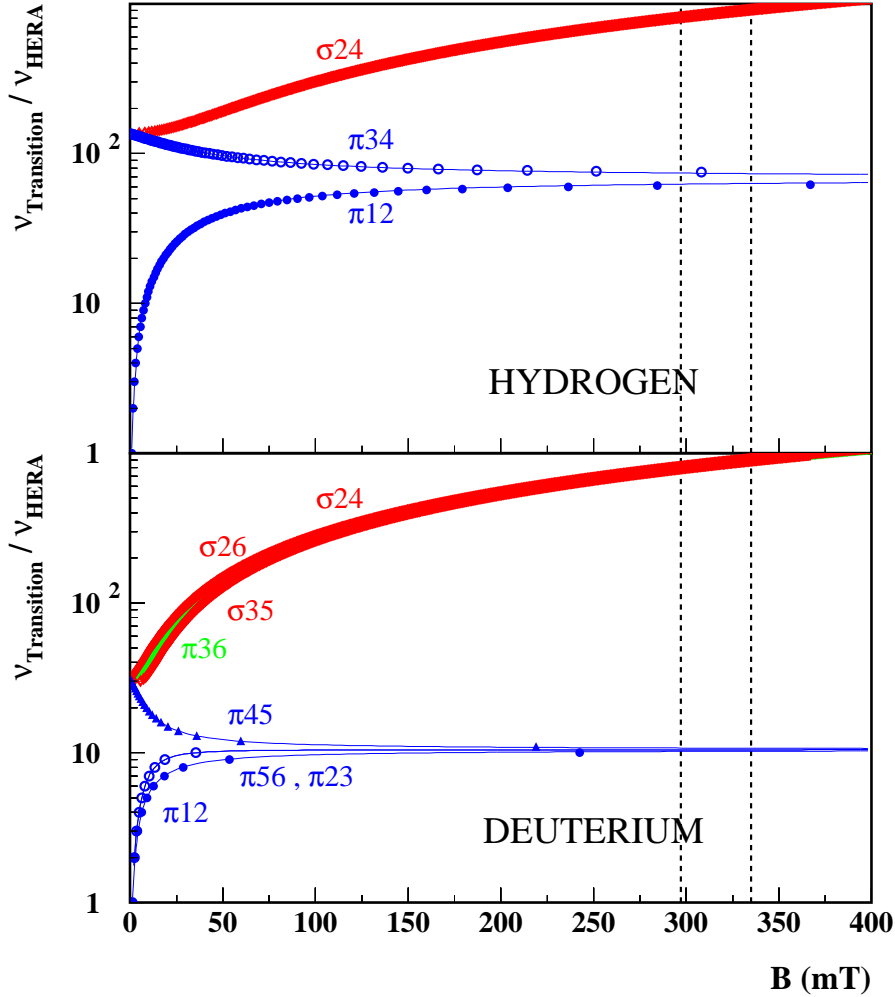} 
\end{center}
\caption{Possible beam-induced nuclear depolarizing resonances in the 
HERMES target. 
The frequency difference between pairs of hyperfine 
states whose transitions lead to nuclear depolarization are plotted as a 
function of the holding field. The frequency values are normalized 
to $\nu_{HERA}=10.42\,$MHz. 
The symbols representing the resonance conditions
are clearly distinguishable for the $\pi$ transitions, while they overlap 
with each other in the case of $\sigma$ and $\pi\ket{3}\leftrightarrow\ket{6}$
transitions,  
which are separated by a difference in the intensity of the holding field
$\Delta B_T$ of only 0.37\,mT.
The dashed lines represent the working fields of the transversely (left) and 
longitudinally (right) polarized targets.}
{\label{fig:reson_tot}}
\end{figure*}

The beam of the HERA electron storage ring is bunched with 
time period $\tau_{bunch}=96.1$\,ns between two adjacent bunches.
Typically, most of the 220 RF buckets are filled, so that the 
bunch frequency is
given by:
% \begin{eqnarray} \nonumber
\begin{equation}
\nu_{bunch} = \frac{1}{\tau_{bunch}} = 10.42  \hspace{1mm}\mathrm{MHz}\,.
\end{equation}
% \end{eqnarray}
A large number
of harmonics contribute to the induced magnetic high frequency field close to
the HERA beam, because the bunches are very short ($\sigma=30$\,ps). 
The transition 
frequency between two hyperfine states $\ket{i}$ and $\ket{j}$ is given by:
\begin{equation}
\nu_{ij} = \frac{E_i(B)-E_j(B)}{h}\,.
\end{equation}
If $\nu_{ij}$ matches one of these harmonics at a certain holding field $B$, 
resonant depolarization occurs.
The location of the beam-induced resonances is shown in 
Fig.~\ref{fig:reson_tot}.
Bunch field induced depolarization at the HERMES target has
been studied in the cases of hydrogen~\cite{Hermes:beamdepol,kol98} and 
deuterium~\cite{Baumgarten:thesis2} for the longitudinal orientation of
the target holding field (before 2001).
During this time the observed $\pi$ ($\Delta m = \pm 1$) transitions were 
sufficiently well separated so that at the working field (335 mT) no beam
induced depolarization could be found. In the case of the transverse field
that has been in use since October 2001, both $\sigma$ ($\Delta m = 0$) 
and $\pi$ ($\Delta m = \pm 1$) transitions are allowed because the oscillating
field surrounding the HERA beam has components both parallel and
perpendicular to the static holding field.
The spacing between two adjacent resonances
is narrower than in the longitudinal case (only 0.37\,mT difference in
the magnetic holding field at $B = 0.3$\,T), so a better homogeneity in the
holding field is needed (see Fig.~\ref{fig:reson_tot}).

Thus a transverse
field value for which beam resonant depolarization can be 
completely suppressed does not exist. A pair of additional correction coils
mounted closely around the storage cell to improve the field 
homogeneity was installed for the 2003 
running to further reduce resonant depolarization.
The results will be presented in a forthcoming paper.

%%%%%%%%%%%%%%%%%%%%%%%%%%%%%%%%%%%%%%%%%%%%%%%%%%%%%%%%%%%%%%%%%%%%%%%%%
%\input{polarization_paolo.tex}

\section{Target polarization}

%Because the systematic uncertainty of any polarized data analysis
%is directly affected by the uncertainty on the target polarization,
%a great effort has been applied to maximize its accuracy.
The average polarization $P_T$ of the nucleons in the target is a 
fundamental parameter of the HERMES experiment.
The problem is to relate the measured atomic polarizations 
in the sample beam with the density-weighted average polarization of the 
nuclei in the cell. The determination of the target polarization involves 
several terms that mainly account for two effects:
\begin{itemize}
\item{Presence of molecules in the cell, both unpolarized from the residual
gas of the scattering chamber and/or directly injected from the ABS, and 
polarized molelcules 
coming from recombination of polarized atoms on the cell surface.}
\item{Relaxation and recombination effects in the storage cell leading to an
inhomogeneous distribution of the atomic fraction and of the atomic
polarization.}
\end{itemize}

The average target polarization $P_T$ as seen by the electron 
beam is given by the following equation:
\begin{equation}
P_T=\alpha_0\alpha_r P_a + \alpha_0(1-\alpha_r) \beta P_a\,,
\label{eqn:pol}
\end{equation}
where $\alpha_0$ represents the initial fraction of nucleons in atoms 
in the gas injected by the ABS 
(or atomic fraction in the absence of recombination),
$\alpha_r$ denotes the fraction of nucleons in atoms
that have survived recombination, and
% during their way through the storage cell,
$\beta = P_m/P_a$ represents the ratio of the nuclear polarization 
of the molecules $P_m$ produced by recombination 
relative to the nuclear polarization of the atoms
$P_a$. The knowledge of $\beta$ for the experimental conditions at the 
HERMES experiment will be discussed in Sec.~\ref{sec:beta}.

%The average polarization $P_T$ of nucleons in the target can be calculated
%using the results of the measurements of the TGA and BRP. $P_T$ has 
%contributions for polarized atoms ($P_a$) and polarized molecules if two
%polarized atoms recombine to a molecule. Nucleons that enter the storage cell
%as molecules are unpolarized. If $n_a$ is the number of atoms inside the 
%storage cell and $n_m$ is the number of molecules, the fraction of nucleons
%in the atomic state (also called the degree of dissociation) $\alpha$ is
%defined by:
%
%\begin{equation}
%\alpha=\frac{n_a}{n_a+2n_m}
%\end{equation}.
%
%With $\alpha_0$ representing the initial fraction of nucleons in atoms as
%injected by the ABS (or atomic fraction in absence of recombination)
%and $\alpha_r$ denoting the fraction of nucleons in atoms
%surviving recombination during their way through the storage cell,
%$P_T$ can be calculated by:
%
%\begin{equation}
%P_T=\alpha_0\alpha_rP_a+\alpha_0(1-\alpha_r)\beta P_a
%\label{eqn:pol}
%\end{equation}
%
%where $\beta = P_m/Pa$ stands for the ratio of the polarization 
%of the molecules 
%produced by recombination $P_m$ relative to the polarization of the atoms
%$P_a$. The knowledge of $\beta$ for the experimental conditions at the 
%HERMES experiment will be discussed in Par.~\ref{sec:beta}.
The values for $\alpha_0$, $\alpha_r$ and $P_a$ are calculated using the
measurements of the TGA and BRP combined with various
calibrations. As both detectors measure the properties of a
sample of the target gas from the center of the target cell, 
it is necessary to relate the results to a 
corresponding value averaged along the storage cell.
This is achieved by evaluating 
the sampling corrections~\cite{molflow}:
\begin{eqnarray} \label{sc:alpha}
c_{\alpha} &\equiv& \frac{\alpha_r}{\alpha_r^{TGA}}\,, \\
c_P &\equiv& \frac{P_a}{P_a^{BRP}}\,.
\label{sc:pol}
\end{eqnarray}
The sampling corrections and their systematic uncertainties are functions of
the measured values of $\alpha_r^{TGA}$ and $P_a^{BRP}$, respectively. The
functions depend on the geometry of the storage cell, its surface properties
and the detectors' acceptances. They are calculated with the help of 
Monte Carlo 
simulations of the stochastic motion of particles in the storage cell, where the
histories of these particles are stored and analyzed. Another possibility is 
to 
calculate sampling corrections analytically by describing the diffusion
process of the target gas inside the storage cell with the use of a
one-dimensional diffusion equation. The issue was addressed in
Ref.~\cite{molflow} and will be discussed here in Sec.~\ref{sec:sc}.

The various quantities entering the calculation of the
average polarization are summarized in Tab.~\ref{pol:terms}.
\begin{table*}
\begin{center}
\begin{tabular}{|c|c|}
\hline
Quantity  &Meaning \\
\hline\hline
$P_T$ & density-averaged nuclear polarization in the cell\\
\hline
$\alpha_0$ & atomic fraction in the absence of recombination within the cell\\
\hline
$\alpha_r = c_\alpha\alpha_r^{TGA}$
&fraction of atoms surviving recombination in the cell\\
\hline
 $c_\alpha$ & sampling correction for the atomic fraction\\
\hline
 $\alpha_r^{TGA}$ & measured atomic fraction in the sample beam\\
% \hline
% $(1-\alpha_r)$  &fraction of atoms recombined in the cell\\
\hline
$P_a =c_P P_a^{BRP}$
&nuclear polarization of atoms in the cell\\
\hline
 $c_P$ & sampling correction for the atomic polarizations\\
\hline
 $P_a^{BRP}$ & measured nuclear polarization of atoms in the sample beam\\
\hline
$\beta = P_m/P_a$ & relative nuclear polarization of recombined atoms\\
\hline
\end{tabular}
\end{center}
\caption{\label{pol:terms}
{Meaning of the quantities contained in eq.~(\protect\ref{eqn:pol}).}}
\end{table*}

\subsection{Determination of the quantities $\alpha_r$ and $\alpha_0$}
\label{section:TGAmeas}

In order to calculate the target polarization following eq.~(\ref{eqn:pol}), 
it is necessary to separate 
the atomic fraction $\alpha_0$ that would be seen by the HERA electron beam 
if no cell wall recombination were to occur and the fraction $\alpha_r$ 
of the initially injected atoms that survive wall recombination,
as seen by the HERA electron beam.
The measured quantity $\alpha^{TGA}$ can also be factored in two
terms and can be written as (see eq.~(\ref{alpha_TGA})):
\begin{equation}
\alpha^{TGA}=\frac{\phi_a}{\phi_a+\phi_m} = \alpha_0^{TGA}\alpha_r^{TGA}\,.
\label{alpha_TGA_2}
\end{equation}
Since dissociative ionization can be neglected~\cite{Hermes:TGA}, 
the only source of
atomic flux $\phi_a$ into the gas analyzer is that originating from
the atomic beam source. The molecular flux $\phi_m$ 
may come from three different sources:
\begin{itemize}
\item Undissociated molecules ballistically injected by the ABS. 
Since the efficiency of the dissociator is around 80$\,\%$, a significant
number of molecules pass through the nozzle towards the polarizing stage,
affected by neither the sextupole magnets nor the transition units. 
Although the majority of these molecules hit the vacuum chamber 
wall and are pumped
 away, a small number enter 
the storage cell via the injection tube. These unpolarized molecules
have a triangular density distribution along the cell axis, 
with the maximum density at the center of the cell.  
\item Rest gas in the target chamber. These molecules originate mainly 
from 
the recombination of atoms that have left the storage cell and have 
thermalized on the vacuum chamber walls. 
This molecular component is clearly unpolarized and results in a uniform
density component inside the storage cell. 
\item Atoms that recombine into molecules while in the cell. 
As explained in Sec.~\ref{sec:cell}, the cell material and temperature
 are optimized in order to limit the depolarization and 
recombination effects in the target atomic 
gas. Nevertheless, atoms can stick on the cell surface and recombine,
and the resulting molecules may carry a residual nuclear polarization. 
\end{itemize}
The normalized total molecular flow rate $\phi_m$ of atoms into
the TGA ionizer can be 
decomposed in the following way:
\begin{equation}
\phi_m=\phi_{bal}+\phi_{rg}+\phi_{r}\,,
\label{phi_rec}
\end{equation}
where $\phi_{bal}$, $\phi_{rg}$ and $\phi_{r}$ represent the ballistic, 
the rest gas and the recombination components, respectively.
To separate between polarized and unpolarized molecules, we rewrite
%one expresses the
%fraction of atoms surviving recombination $\alpha_r^{TGA}$ and the initial
%atomic fraction $\alpha_0^{TGA}$ as measured by the TGA by rewriting 
the two factors in eq.~(\ref{alpha_TGA_2}) using the three contributions
appearing in eq.~(\ref{phi_rec}):
\begin{eqnarray}
\alpha_r^{TGA} &\equiv& \frac{\phi_{a}}{\phi_{a}+\phi_{r}}\,,
\label{eq:alpha_r_TGA} \\
\alpha_0^{TGA} &\equiv&
\frac{\phi_{a}+\phi_{r}}{\phi_{a}+\phi_{r}+\phi_{bal}+\phi_{rg}}\,.
\label{eq:alpha_0_TGA}
\end{eqnarray}

The independent measurement of each of the three components of the total molecular flow rate 
is accomplished by means of routine {\it target calibrations}, 
which are usually performed between two HERA electron fills
when the HERMES spectrometer is not taking data. 
The rest gas contribution $\phi_{rg}$ to the molecular flux is linearly 
proportional to the pressure in the target chamber.  The constant
of proportionality is determined by injecting into the target chamber
increasing quantities of hydrogen (deuterium) molecules with a controlled
flow system and by measuring the corresponding TGA count rates.
The ballistic flux contribution $\phi_{bal}$
can be measured by varying the 
atomic flux of the injected beam while keeping the molecular flux
constant.  The atomic flux is changed by using the hyperfine
transitions to select either 1 or 2 (or 3 in the case of deuterium) 
hyperfine states.  Since the recombination probability inside the cell
was found to be independent of the injected beam 
intensity~\cite{Hermes:recomb}, the quantity
$\phi_{bal}$ can be extracted. Eq.~(\ref{phi_rec}) is then solved for the
remaining molecular flux $\phi_r$.
After the molecular contributions $\phi_{rg}$, $\phi_{bal}$ and $\phi_r$ are
extracted, one can determine $\alpha_r^{TGA}$ using 
eq.~(\ref{eq:alpha_r_TGA}) and thus $\alpha_r$ using eq.~(\ref{sc:alpha})
with the sampling correction $c_\alpha$ determined as described
in Sec.~\ref{sec:sc}.
A detailed description of how these calibrations
are performed was given in Ref.~\cite{Hermes:TGA}.

The factor $\alpha_0$ is defined in terms of the nucleon thicknesses
$t_{a,r,bal,rg}$ in the cell seen by the HERA beam:  
\begin{equation}
\alpha_0 \equiv \frac{t_a+t_r}{t_a + t_r + t_{bal}+t_{rg}}\,.
\end{equation}
It can be determined without reference to additional sampling
corrections.  All the $\phi_i$ terms in eq.~(\ref{eq:alpha_0_TGA})
represent nucleon fluxes.  Let us define a proportionality
constant $c$ representing the TGA response to atoms through the
relationship $\phi_a = c t_a$. The fluxes $\phi_r$, $\phi_{bal}$
and $\phi_{rg}$ are suppressed by a factor of $\sqrt{2}$ due to
the smaller molecular velocity, leading to $\phi_{r} = c
t_{r}/\sqrt{2}$ and $\phi_{bal} = c t_{bal}/\sqrt{2}$.
Furthermore, in the limiting case of small cell wall recombination, 
the densities of atoms and molecules producing $t_a$, $t_{bal}$ and $t_r$ 
are triangular. In contrast, the rest gas 
producing $t_{rg}$ has a uniform density along the cell,
resulting in $\phi_{rg} = c t_{rg}/(2 \sqrt{2})$. 
If the difference between the conductances of the atoms and
molecules is neglected so that they have the same distributions
along the cell, the above considerations can be combined to give
\begin{equation}
\alpha_0  = \frac{\phi_a + \sqrt{2}\phi_r}
{\phi_a + \sqrt{2}(\phi_r + \phi_{bal} + 2\phi_{rg})}\,.
\label{eqn:simplea0}
\end{equation}
However, this difference in conductances can be taken into
account~\cite{Baumgarten:thesis2,Reggiani:thesis}, 
leading to different distributions along the cell
and a more complicated expression:
\begin{equation}
\alpha_0 =
\frac{d(\phi_a+\phi_r)}
{d(\phi_a+\phi_r)
+\sqrt{2}(\phi_{bal}+2\phi_{rg})}\,,
\label{eqn:a0}
\end{equation}
where $d=\sqrt{2}/(1+(\sqrt{2}-1)\alpha_r)$.

\subsection{Relative polarization $\beta$ of nuclei in recombined atoms }
\label{sec:beta}

As mentioned in the previous sections,
some of the polarized hydrogen or deuterium atoms may recombine
in the storage cell. 
A direct measurement of the remnant polarization contained in the
molecules is not possible at HERMES, as the Breit-Rabi 
polarimeter can measure the atomic polarizations only.
Atomic recombination is a  surface-mediated 
temperature-dependent process~\cite{Hermes:recomb,Baumgarten:thesis2}. 
Hence measurements of the nuclear polarization of recombined atoms 
using storage cells of various materials and/or with various coatings
~\cite{vdb_paper,wis_paper} are not directly applicable to the HERMES case.
Without any information about the polarization $\beta$ of
nucleons in molecules (or molecular polarization in short) one has 
to allow in eq.~(\ref{eqn:pol}) the conservative range $0\le\beta\le1$.
A set of data taken in 1997 at a higher cell temperature 
(260\,K instead of the nominal 100\,K) was used to measure
$\beta$ for hydrogen at that temperature, resulting 
in $\beta^{260\,K}=0.68\pm0.09_{stat}\pm0.06_{syst}$~\cite{Hermes:beta}.
This result can be used as an upper limit for $\beta$ at 100\,K 
($\beta^{100\,K}_{high}$) under the following assumptions.
The main mechanism responsible for recombination at 100\,K and 260\,K is the
same (i.e. the E-R mechanism)~\cite{Hermes:recomb}.
At 100\,K, the recombination probability is smaller
than at 260\,K, because the atoms impinging from the volume have
less kinetic energy to overcome the activation barrier.  
The residence time of chemically adsorbed atoms on 
the surface at 100\,K is greater than at 260\,K, 
and their possible polarization cannot exceed 
the value measured at 260\,K. This results in $\beta^{100\,K}_{high}=0.83$.

The lower limit for $\beta^{100K}$ ($\beta_{low}^{100K}$)
can be derived 
from a simple argument.
By assuming that the nucleon spins are not affected by the
recombination process,
the nuclear polarization of the molecule at the time of 
its formation ($P^0_m$) 
can be evaluated by taking the average value of the 
polarization of an atom coming from
the volume ($P{_a}$) and one resident on the surface ($P{_s}$).
The theoretical lower limit for $\beta^{100K}$ can then be inferred
by supposing that atoms sticking on the cell surface 
are totally depolarized ($P_s^{100K,low}=0$) and applying the equation:
\begin{eqnarray} \nonumber
\beta^{100K}_{low}&=&
\frac{P_{m}^{100K,low}}{P_a^{100K}} \\
&=&\frac{P_a^{100K}+P_s^{100K,low}}{2}\frac{1}{P_a^{100K}}\,.
\end{eqnarray}
The result is $\beta_{low}^{100K} = 0.5$. 
A factor of 0.9 takes into account the depolarization of molecules 
while colliding with the walls~\cite{Hermes:beta,wis_paper}, ending in
 $\beta_{low}^{100K} = 0.45$.
Hence, the range of $\beta^{100K}$ can be limited to
\begin{equation}
0.45\le\beta^{100K}\le 0.83\,.
\label{eq:beta_100_range}
\end{equation}
The corresponding value and uncertainty of $\beta^{100K}$ 
are therefore given by:
\begin{equation}
\beta^{100K}=0.64\pm0.19\,.
\label{eq:beta_100}
\end{equation}

\subsection{Injected atomic polarizations}
\label{sec:injectedpol}

An important parameter for the evaluation of the target polarization
is the
injected atomic polarization $P_a^{inj}$ (i.e. the nuclear polarization 
of atoms in the absence of depolarization),
as it enables the separation of the various spin-relaxation effects.
The injected polarization for each ABS operating mode 
(Tab.~\ref{tab:injections}) can be calculated  
if the transmission probabilities of the sextupole system and the efficiencies 
of the adiabatic transition units are known. 
The sextupole transmission probabilities for hydrogen 
and deuterium have been calculated using a Monte Carlo simulation and 
the results are given in Ref.~\cite{Hermes:ABS}. 
The injected polarization is fairly insensitive to these values.
The efficiencies of the transitions of the ABS 
can be extracted by measuring the magnetic field dependence 
of the spin relaxation processes and by fitting the data with a theoretical 
model described in 
Sec.~\ref{sec:maseq} and given by eq.~(\ref{eq:master}).
For normal operation, the important ABS injection 
modes are those having a large positive or
 negative nuclear 
polarization, $P_{z+}$ and $P_{z-}$ respectively, and, in the case of deuterium, 
the two additional positive and negative tensor polarization 
($P_{zz+}$ and $P_{zz-}$) modes (see Tab.~\ref{tab:injections}).
\begin{table}[b!]
\begin{center}
\begin{tabular}{|c|c|c|}
\hline
\rule[-0.3cm]{0cm}{0.8cm}
Mode  & Hydrogen & Deuterium\\
\hline\hline
\rule[-0.2cm]{0cm}{0.6cm}
$P_{z+}^{inj}$&$ 97.26\pm 0.05$&$ 87.78\pm 0.07$\\
\hline
\rule[-0.2cm]{0cm}{0.6cm}
$P_{z-}^{inj}$&$ -97.38\pm 0.07$&$-89.50\pm 0.10$\\
\hline
\rule[-0.2cm]{0cm}{0.6cm}
$P_{zz+}^{inj}$& - &$ 96.22 \pm 0.38$\\
\hline
\rule[-0.2cm]{0cm}{0.6cm}
$P_{zz-}^{inj}$& - &$-177.30 \pm 0.61$\\
\hline
\end{tabular}
\end{center}
\caption{\label{tab:inj_pol}
{Estimated polarizations for the hydrogen and 
deuterium atomic beams injected by the ABS into the target. 
All values are multiplied by 100.  The quoted uncertainties 
take only the dominant uncertainties of the ABS efficiencies 
into account. The values refer to the polarization of injected gas
with the 0.3\,T holding field. Note that, due to the standard way 
of defining it,
the maximum achievable negative tensor polarization is -200.}}
\end{table}

For hydrogen, only two transitions 
after the sextupole system are used: the SFT 2-4 for $P_{z+}$ and 
WFT 1-3 for $P_{z-}$. Therefore, the injected polarizations are directly 
given by the transition efficiencies whose values have been extracted
in Ref.~\cite{Baumgarten:thesis2}
with the technique mentioned above, and cross-checked using a more 
direct method based on the symmetry of the spin relaxation processes 
for the two injection modes.
In the case of deuterium, several transition units are used to provide the 
four polarizations, some of which are located between the two 
sextupole subsystems. 
Hence, the injected polarizations have to be calculated
 taking into account the transition efficiencies as well as the 
transmission probabilities for atoms whose hyperfine state gets changed 
between the two subsystems.

The calculated polarization for the hydrogen and deuterium beams
injected into the target are 
summarized in Tab.~\ref{tab:inj_pol}.   
The degree of nuclear polarization injected by the ABS when running 
with deuterium is substantially lower than for 
hydrogen. The reason is that, due 
to the higher number of hyperfine states, the deuterium nuclear polarization 
can be achieved only by using the transition units located between the 
two magnetic subsystems, resulting in an incomplete separation of states.

\subsection{Sampling Corrections}
\label{sec:sc}

The sampling correction functions $c_{\alpha}(\alpha_r^{TGA})$ and
$c_P(P_a^{BRP})$ depend strongly on the distribution of recombination and wall
relaxation probabilities per wall collision along the storage cell.
%This becomes evident if one compares the histories of a particle entering
%the TGA and another particle still inside the storage cell: they both 
%experience a different amount of wall collisions at different places of 
%the storage cell geometry. This comparison is of special interest because
%the difference in these particle history leads to the sampling correction
%$c_{\alpha}$.
For a freshly produced storage cell that has not yet been 
exposed to the HERA beam, one can assume that 
the recombination and wall relaxation probabilities are
constant over the entire surface of the storage cell, or at least that 
microscopic inhomogeneities due to various types of surface sites are
equally distributed. 
%During such occasions especially the values of $\alpha_r^{TGA}$ and $P_e$
%were affected. Another experimental finding is that during injection the 
%storage cell temperature increases and $P_e$ decreases due to bunch 
%field depolarization~\cite{kol98}.
%During the periods when $\alpha_r^{TGA}$ and $P_e^{BRP}$ have been affected,
%increased synchrotron radiation and/or charged rates have been measured
%by the HERMES hodoscopes and the vacuum pressure increases, which indicates
%an unusual rise of the heat load at certain parts of the beam pipe.
Possible mechanisms by which the HERA beam could affect the surface properties
of the storage cell include:
\begin{itemize}
\item{desorption of the surface coating by synchrotron radiation;}
\item{bombardment of the storage cell walls with charged particles
leading to radiation damage;}
\item{chemical reactions of ionized hydrogen or deuterium radicals with
the cell surface;}
\item{redistribution 
of the ice layer on the storage cell by RF heating
due to the wake fields;}
\item{contamination of the storage cell surface by sputtered material from
the collimator upstream of the target.}
\end{itemize}
Under normal operating conditions
no influence of the HERA beam on
the target performance has been observed. Nevertheless certain distinct
incidents affecting the cell have been observed, when the beam tune was not
optimal or the HERA beam was accidentally lost near the HERMES target region.
Since the effect of the HERA beam on the surface properties is poorly known,
every realistic distribution of recombination and wall relaxation probabilities
has to be considered when evaluating the sampling corrections.
It is therefore convenient to distinguish certain 
scenarios of special interest.
\begin{itemize}
\item{{\bf{Homogeneous cell (HC):}} the natural 
assumption for a freshly produced cell not yet
exposed to the HERA beam. The calculated sampling corrections 
set an upper limit for $\alpha_r$ and $P_a$ for any given values 
of $\alpha_r^{TGA}$ and $P_a^{BRP}$, respectively.}
\item{{\bf Homogeneous beam tube (HBT):} 
the main cell tube is homogeneously affected by the beam, while the side tubes 
are unaffected.}
\item{{\bf Inhomogeneous beam tube (IBT):} The main cell tube is 
inhomogeneously affected by the beam, while the side tubes are unchanged. 
This scenario does not 
lead to a single sampling correction, but it can be used to calculate 
a lower limit for $\alpha_r$ or $P_a$ for any given value of 
$\alpha_r^{TGA}$ or $P_a^{BRP}$, respectively.}
\end{itemize}

A model for calculating the sampling corrections was developed in 
Ref.~\cite{Baumgarten:thesis2}. It is based on a parameter
$\gamma_r (\gamma_d)$, the recombination (spin-flip) probability
during a wall collision. If these probabilities were constant and equal
for the beam and sample tubes, the sampling corrections would depend only 
on their geometry. This is actually the case for a freshly installed cell,
whose Drifilm coating is known to be intact and uniform.
In all scenarios it is assumed that the relation 
$\gamma_r^{ST}\le\gamma_r^{BT}$ holds at any time 
for the recombination probabilities in the
sample tube, $\gamma_r^{ST}$, and beam tube, $\gamma_r^{BT}$. 
The validity of this hypothesis 
is proven by the data~\cite{Baumgarten:thesis2,Reggiani:thesis}.
The numerical calculations of the sampling corrections have 
been performed using a molecular flow simulation, complemented by
analytic calculations based on the one-dimensional diffusion 
equation~\cite{Baumgarten:thesis2}. 

\begin{figure*}
\begin{center}
\includegraphics[width=\textwidth]{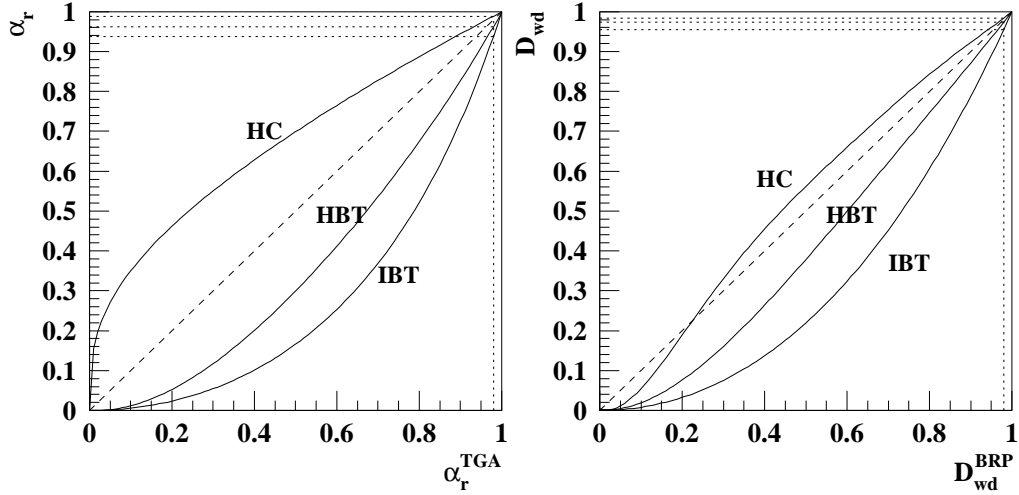}
\end{center}
\caption{Simulated correlation between the values measured by the TGA or
BRP detectors and their corresponding density-weighted average values in the
storage cell. The three solid curves represent the three scenarios 
for the state of the cell surface: homogeneous cell (HC), 
homogeneous beam tube (HBT), and inhomogeneous beam tube (IBT).
The left figure shows the range for recombination and
the right picture shows the range for wall depolarization.
Acceptable uncertainties of $\alpha_r$ and $\depol_{WD}$ 
(differences among the three horizontal dotted lines)
can be achieved only with $\alpha_r^{TGA}$ and $\depol_{WD}^{BRP}$ 
close to unity (vertical dotted lines). The corresponding uncertainty
grows rapidly with increasing recombination or depolarization.}
\label{fig:sampling}
\end{figure*}

The sampling correction $c_P$ must
be subdivided into two terms, one for the wall collisions 
and a second term for spin exchange relaxation, as the two processes
must be treated in different ways. This can be done by defining 
the depolarization factors $\depol_{wd}$ (wall collisions) 
and $\depol_{se}$ (spin exchange relaxation) so that
for small spin relaxation, one has:
\begin{equation}
c_P\equiv\frac{P_{a}}{P_a^{BRP}}\simeq\frac{\depol_{wd}\,\depol_{se}\,P_a^{inj}}
{\depol^{BRP}_{wd}\,\depol^{BRP}_{se}\,P_a^{inj}}\,.
\label{eq:P_decompose}
\end{equation}
As the spin exchange relaxation occurs in the gas phase, and for
the typical values of $\alpha_r>0.9$ the atomic density distribution
is little affected by any change of the cell surface properties,  
the relation between $\depol_{se}$ and $\depol^{BRP}_{se}$ is 
similar for all
of the listed scenarios. Moreover, since calculations have 
shown that $\depol_{se}\simeq \depol^{BRP}_{se}$~\cite{Baumgarten:thesis2}, 
one can conclude that the sampling correction $c_P$ is determined by 
the wall depolarization only:
\begin{equation} \label{eq:seapprox}
c_P=\frac{P_a}{P_a^{BRP}}\simeq \frac{\depol_{wd}}{\depol_{wd}^{BRP}}\,.
\end{equation}
As an example, in Fig.~\ref{fig:sampling} the effect of 
the calculated sampling corrections for
the large-aperture  cell used in the 1996-97 running with hydrogen is shown.
During normal operation, the values of $\alpha_r^{TGA}$ and $\depol_{wd}^{BRP}$ 
are found to be almost always well above $0.9$, so that local linear 
approximations may be fitted to the calculated sampling corrections:
\begin{eqnarray}
\alpha_r&=&\bar a+(1-\bar a)\,\alpha_r^{TGA} \\
&=&\alpha_r^{TGA}+\,\bar a\,(1-\alpha_r^{TGA})\,,
\label{eq:alpha_samp_lin} \\
% \end{eqnarray}
% \begin{eqnarray}
\depol_{wd}&=&\bar b+(1-\bar b)\,\depol_{wd}^{BRP} \\
&=& \depol_{wd}^{BRP}+\,\bar b\,(1-\depol_{wd}^{BRP})\,.
\label{eq:pol_samp_lin}
\end{eqnarray}
The values of the parameters $\bar a$ for recombination and $\bar b$
for depolarization are fitted to the average between the two extreme 
scenarios shown in Fig.~\ref{fig:sampling}, and
their systematic uncertainties $\Delta \bar a$ and $\Delta \bar b$ 
are taken to be the differences found when fitting the extremes.
% (see the dotted lines in Fig.~\ref{fig:sampling}).

\subsection{Evaluation of the systematic uncertainty in $P_T$}

After having determined all the terms of eq.~(\ref{eqn:pol}), 
$P_T$ can be expressed as a function of the measured quantities. 
It is convenient to write the target polarization in the following way:
\begin{equation}
P_T=\alpha_0\,(\alpha_r+(1-\alpha_r)\,\beta)P_a=\alpha_{eff}\cdot P_a\,.
\label{eq:alpha_tot}
\end{equation}
where the term $\alpha_{eff}$ is called {\it{effective}} atomic fraction.

\subsubsection{Uncertainty of the effective atomic fraction $\alpha_{eff}$}

Combining eqs.~(\ref{eq:alpha_tot}), 
(\ref{eq:alpha_samp_lin}), (\ref{eqn:a0}), (\ref{eq:alpha_r_TGA}) 
and (\ref{phi_rec}), $\alpha_{eff}$ can be expressed as
\begin{equation}
\alpha_{eff}={\phi_a+(\bar a\,(1-\beta)+\beta)(\phi_m-\phi_ {bal}-\phi_{rg})\over
\phi_a+\phi_m+(\sqrt{2}-1)\phi_ {bal}+(2\sqrt{2}-1)\phi_{rg}}\,.
\label{eq:alpha_tot_phi}
\end{equation}
% \begin{equation}      % simplified version
% \alpha_{eff}=
% \frac{\phi_a + \sqrt{2}\phi_r}
% {\phi_a + \sqrt{2}(\phi_r + \phi_{bal} + 2\phi_{rg})}
% \left(\beta+(1-\beta)\frac{\phi_a+\bar{a}\phi_r}{\phi_a+\phi_r}\right)\,.
% \label{eq:alpha_tot_phi}
% \end{equation}
The systematic uncertainty of $\alpha_{eff}$ is determined by the 
measured flow rates $\phi_a$, $\phi_m$, $\phi_ {bal}$ and $\phi_{rg}$ 
and their uncertainties as well as the values and uncertainties of 
$\beta$, $\bar a$ and the TGA calibration constant $\kappa$. 
With the matrix $T$ defined by
\begin{equation}
T=\left(\frac{\partial{\alpha_{eff}}}{\partial{\phi_i}},\dots,
\frac{\partial{\alpha_{eff}}}{\partial{\kappa}}\right)\,,
\end{equation}
and the diagonal covariance matrix $C$ of all calibration constants, 
the systematic uncertainty in $\alpha_{eff}$ is given by
\begin{equation}
\Delta_{sys}\alpha_{eff}=\sqrt{TCT^T}\,.
\end{equation}
The matrix $T$ contains 10 independent derivatives, i.e. for $\kappa$, 
$\bar a$, $\beta$ and the 7 additional terms 
related to the ballistic flux and rest gas calibrations
as described in Ref.~\cite{Hermes:TGA}. 
%The uncertainty source connected to the
%last six quantities is collected in a reduced 
%uncertainty $\tilde\Delta\phi_ {bal}$ which does not take into account 
%$c_{rg}$. The remaining derivatives can by calculated from 
%Eq. (\ref{eq:alpha_tot_phi}) by

%\begin{eqnarray}
%\pdif{\alpha_{eff}}{\kappa}=&\left[(\bar a(1-\beta)+\beta-\alpha_{eff})\phi_r
%-\alpha_{eff}\sqrt{2}(2\phi_{rg}+\phi_ {bal})\right]/(\kappa N)\\
%\pdif{\alpha_{eff}}{c_{rg}}&=&-\left[(\bar a(1-\beta)+\beta+(\sqrt{2}-1)\alpha%_{tot})(p_{tc}+\pdif{\phi_ {bal}}{c_{rg}})+\alpha_{eff}\sqrt{2}p_{tc}\right]/N%\\
%\pdif{\alpha_{eff}}{\bar a}&=&(1-\beta)\phi_r/N\\
%\pdif{\alpha_{eff}}{\beta}&=&(1-\bar a)\phi_r/N\\
%\pdif{\alpha_{eff}}{\phi_ {bal}}&=&-\left[\bar a(1-\beta)+\beta+
%(\sqrt{2}-1)\alpha_{eff}\right]/N\\
%\end{eqnarray}
%where $N=\phi_a+\phi_m+(\sqrt{2}-1)\phi_ {bal}+(2\sqrt{2}-1)\phi_{rg}$ 
%is the denominator
%of equation (\ref{eq:alpha_tot_phi}) and $\phi_r$ is given 
%by (\ref{phi_rec}).\\
%The derivative $\pdif{\phi_ {bal}}{c_{rg}}$ and the value 
%of $\tilde\Delta\phi_ {bal}$ are derived from the calibrations described in
%Sec.~\ref{section:TGAmeas}:
%\begin{eqnarray}
%\pdif{\phi_ {bal}}{c_{rg}}&=&-(\phi_a^{(1)}p_{tc}^{(2)}-\phi_a^{(2)}p_{tc}^{(1)})
%/(\phi_a^{(1)}-\phi_a^{(2)})\\
%
%\nonumber
%\tilde\Delta\phi_ {bal}^2&=&[
%(\phi_r^{(2)}\Delta\phi_a^{(1)})^2+
%(\phi_r^{(1)}\Delta\phi_a^{(2)})^2+
%(\phi_a^{(2)}\Delta\phi_m^{(1)})^2+
%(\phi_a^{(1)}\Delta\phi_m^{(2)})^2\\
%&+&(\phi_a^{(2)}c_{rg}\Delta p_{tc}^{(1)})^2+
%(\phi_a^{(1)}c_{rg}\Delta p_{tc}^{(2)})^2
%]/(\phi_a^{(1)}-\phi_a^{(2)})^2.
%\end{eqnarray}

\subsubsection{Uncertainty of the atomic polarizations $P_a$}

Combining eqs.~(\ref{eq:P_decompose}), (\ref{eq:seapprox}) 
and (\ref{eq:pol_samp_lin}), 
% and using the approximation:
% \begin{equation}
% \depol_{se}^{BRP}\simeq\frac{1}{1+\Delta (P_a^{se})^{BRP}}\,,
% \end{equation}
the average nuclear polarization 
$P_a$ of atoms in the HERA beam can be written as
\begin{equation}
P_a= P_a^{BRP}+\bar b\,(\depol_{se}\,P_a^{inj}-P_a^{BRP})
% \simeq P_a^{BRP}+\bar b\left({P_a^{inj}\over 1+\Delta P_a^{se}}-P_a^{BRP}\right)\,.
\label{eq:Pa_calc}
\end{equation}
For the calculation of the systematic uncertainty $\Delta_{sys}\,P_a$, 
the important uncertainty sources to be included 
are $\Delta \bar b$ and $\Delta_{sys}\,P_a^{BRP}$\footnote{The uncertainties 
on $\Delta P_{se}$ and on $P_a^{inj}$ are negligible.}. As a result, 
one finds:
\begin{multline}
% \begin{equation}
\Delta_{sys}P_a=\left((1-\bar b)^2 (\Delta_{sys}P_a^{BRP})^2 
+(1-\depol_{wd})^2\depol_{se}^2 (P_a^{inj})^2\Delta \bar b^2\right)^{1/2}\,.
% \end{equation}
\end{multline}
During the analysis one typically averages over a large dataset, so that 
the statistical uncertainty of the polarization measurement 
$\Delta_{sta}\,P_a$ is negligible.

%%%%%%%%%%%%%%%%%%%%%%%%%%%%%%%%%%%%%%%%%%%%%%%%%%%%%%%%%%%%%%%%%%%%%%%%%%%
%\input{performance.tex}

\section{Performance}
\label{sec:performance}

In the present section an overview is given 
about the target performance in the
running period 1996-2003, by analyzing distinct periods for each
target:
\begin{itemize}
\item{1997 for the longitudinal hydrogen target;}
\item{2000 for the longitudinal deuterium target;}
\item{2002 for the transverse hydrogen target.}
\end{itemize}

\subsection{Longitudinal hydrogen running in 1997}

\begin{figure*}
\begin{center}
\begin{tabular}{c}
\includegraphics[width=\textwidth]{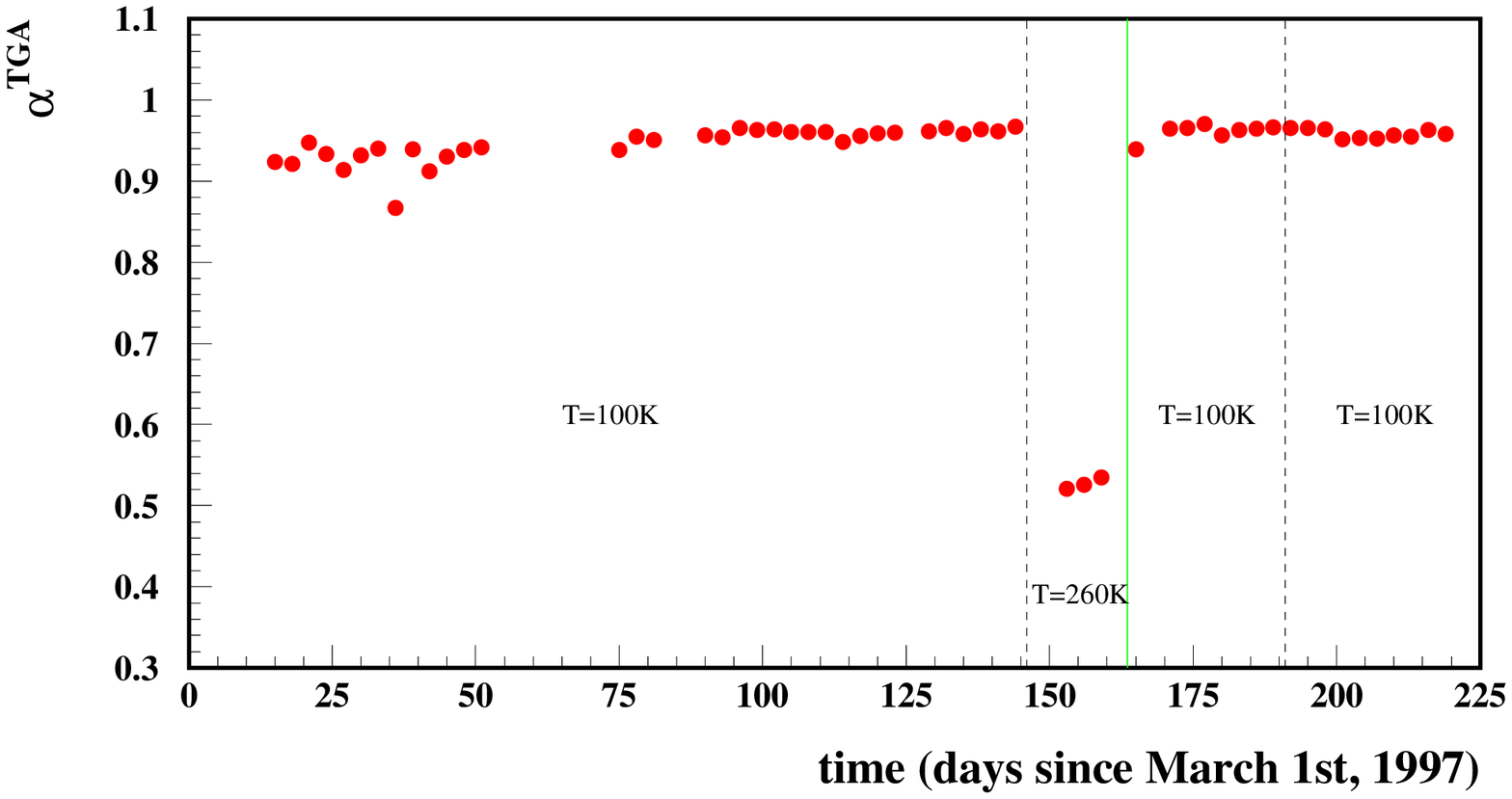}  \\
\includegraphics[width=\textwidth]{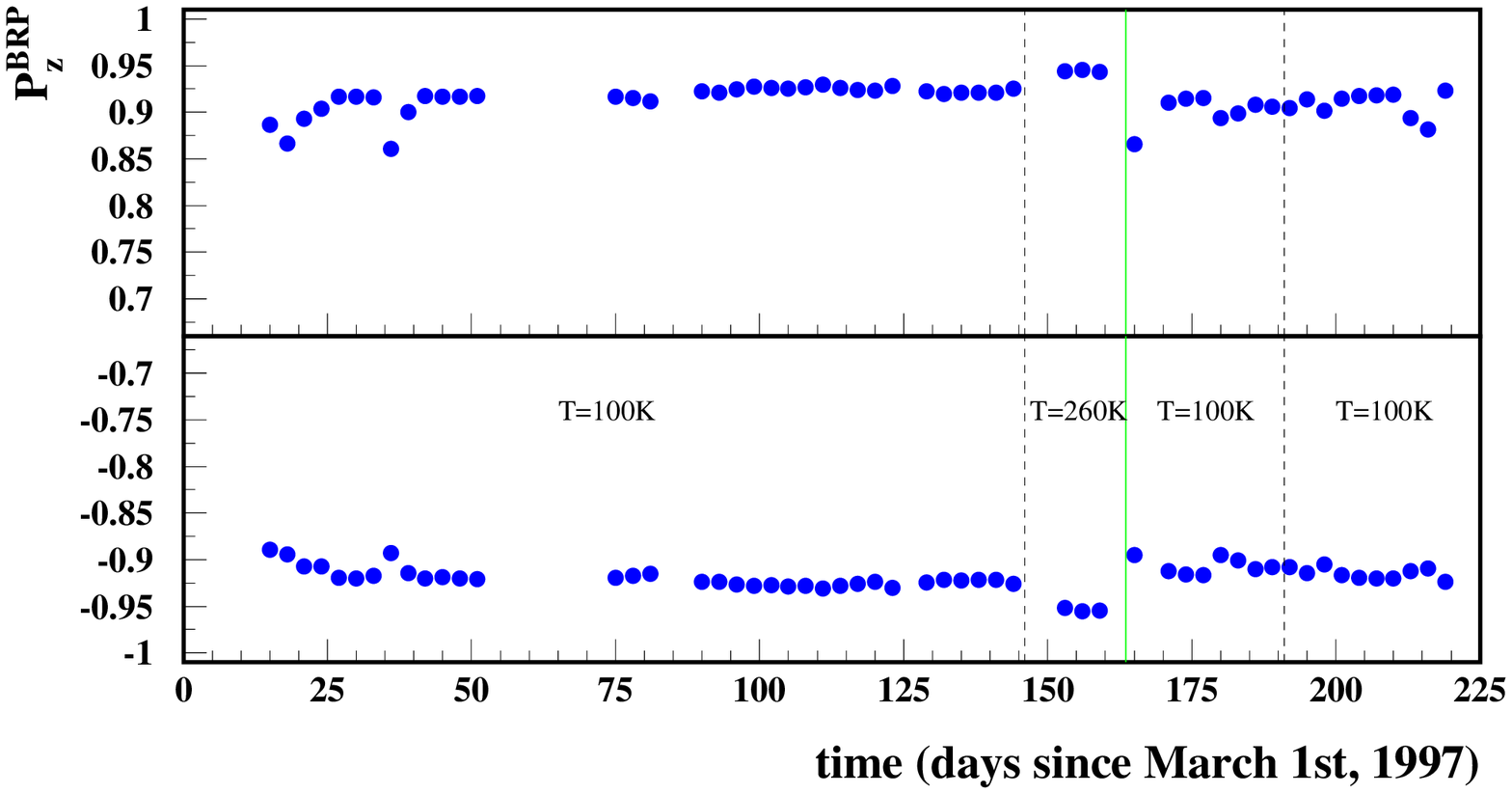} 
\end{tabular}
\end{center}
\caption{The TGA (above) and BRP (below) measurements for the entire
1997 data taking period.  The vertical 
dashed lines indicate HERA beam loss events
which affected the cell surface properties, and the solid line
indicates the replacement of the target cell.}
\label{fig:rawmeas}
\end{figure*}

The upper plot in Fig.~\ref{fig:rawmeas} shows the atomic 
fraction $\alpha^{TGA}$ 
measured by the target gas analyzer during the 1997 running period. 
The nuclear polarization of atoms measured by the BRP is plotted in 
the lower half of Fig.~\ref{fig:rawmeas}. 
Aside from startup problems, the target 
performance was relatively smooth. Unfortunately, a severe HERA beam 
loss close to the HERMES region occurred at a certain point
(left line), resulting in 
a change of the cell surface properties. After this bad event, it was 
decided to collect a set of data at higher temperature ($T_{cell}=260\,$K 
instead of the nominal value $T_{cell}=100\,$K) in order to measure the 
$\beta$ parameter (see Ref.~\cite{Hermes:beta}). At the end of this period, 
the storage cell was exchanged (middle line). Also the second cell 
experienced a beam loss (right line) which produced a small 
hole in the cell wall.
For these reasons, the analysis of the 1997 dataset was
divided into four periods, during each of which the cell surface condition 
was assumed to be constant. For each period, specific studies 
have been performed to limit the uncertainties on $\alpha_r$ and 
$P_a$~\cite{Reggiani:thesis}. 

For the entire running period with hydrogen,
we quote a luminosity-weighted average value for the target 
polarization of $P_{z+}=-P_{z-}=0.852\pm0.033$ and a density
of $7.6\times10^{13}$~nucl./cm$^2$ .
The latter number was derived from a method exploiting 
the density dependence of spin exchange 
collisions~\cite{Baumgarten:thesis2}.
%%%%%%%%%%%%%%%%%%%%%%%%%%%%%%%%%%%%%%%%%%%%%%%%%%%
%%%%%%%%%%%%%%%%%%%

\subsection{Longitudinal deuterium running in 2000}

Over the eight month period of deuterium running in 2000, the HERMES target 
operated under very stable conditions. The atomic beam source, equipped with a 
microwave dissociator, 
experienced only one major failure, which hardly affected data accumulation.
Neither the scattering chamber nor the TGA/BRP vacuum system had to be opened,
 leading to constant amounts of residual gas in the storage cell 
as well as stable efficiencies of the detectors monitoring the target. 
There was no need to replace the storage cell, 
as its performance in terms of atomic 
recombination and nuclear depolarization was always excellent. 
The reason for such good behavior can be found both in a particularly 
fortunate Drifilm coating and in the limited damage caused by a very 
well tuned electron beam, i.e. no unwanted beam losses in the vicinity
of the target.

\begin{figure*}
\begin{center}
\begin{tabular}{c}
\includegraphics[width=0.85\textwidth]{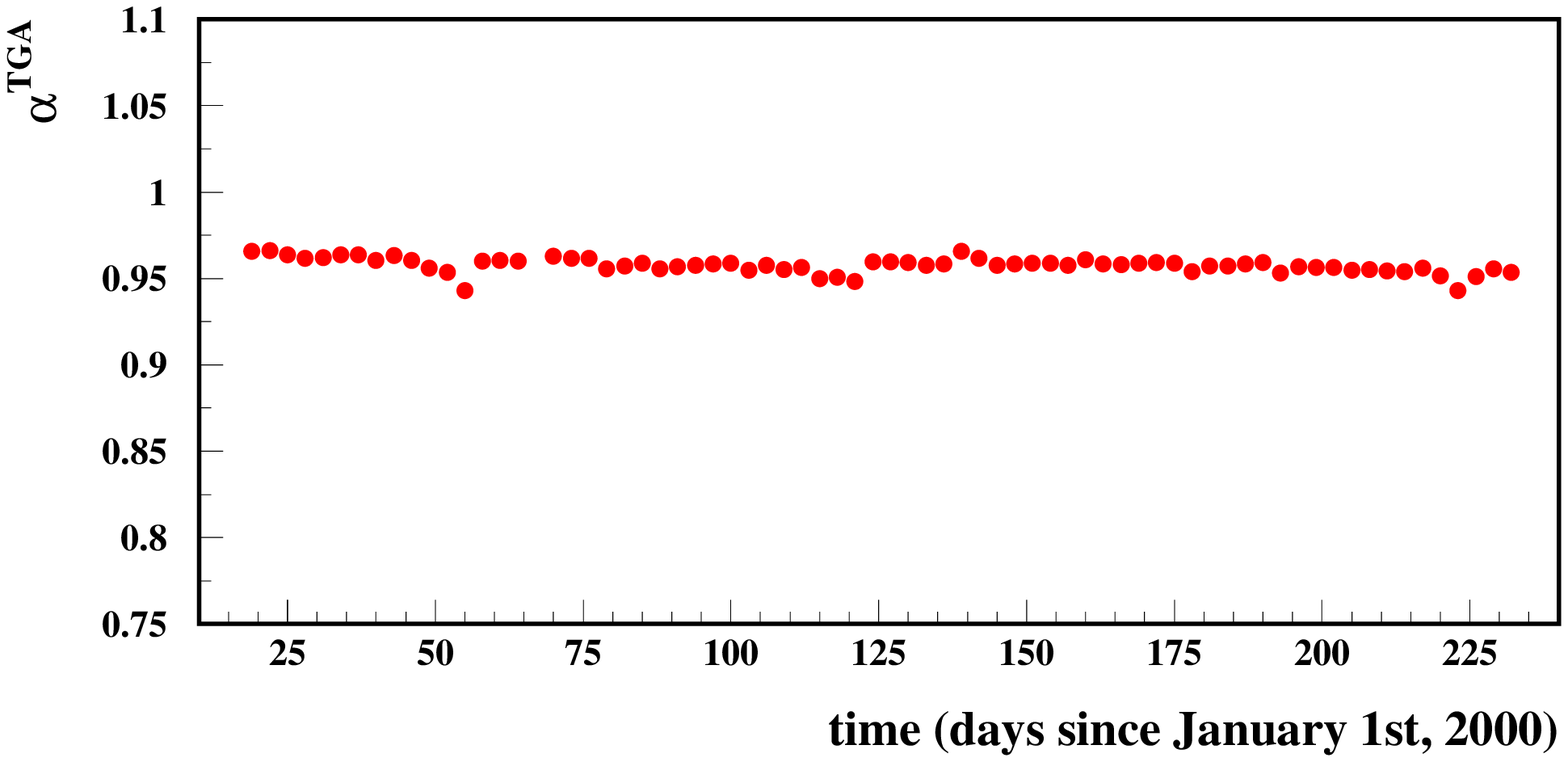} \\
\includegraphics[width=0.85\textwidth]{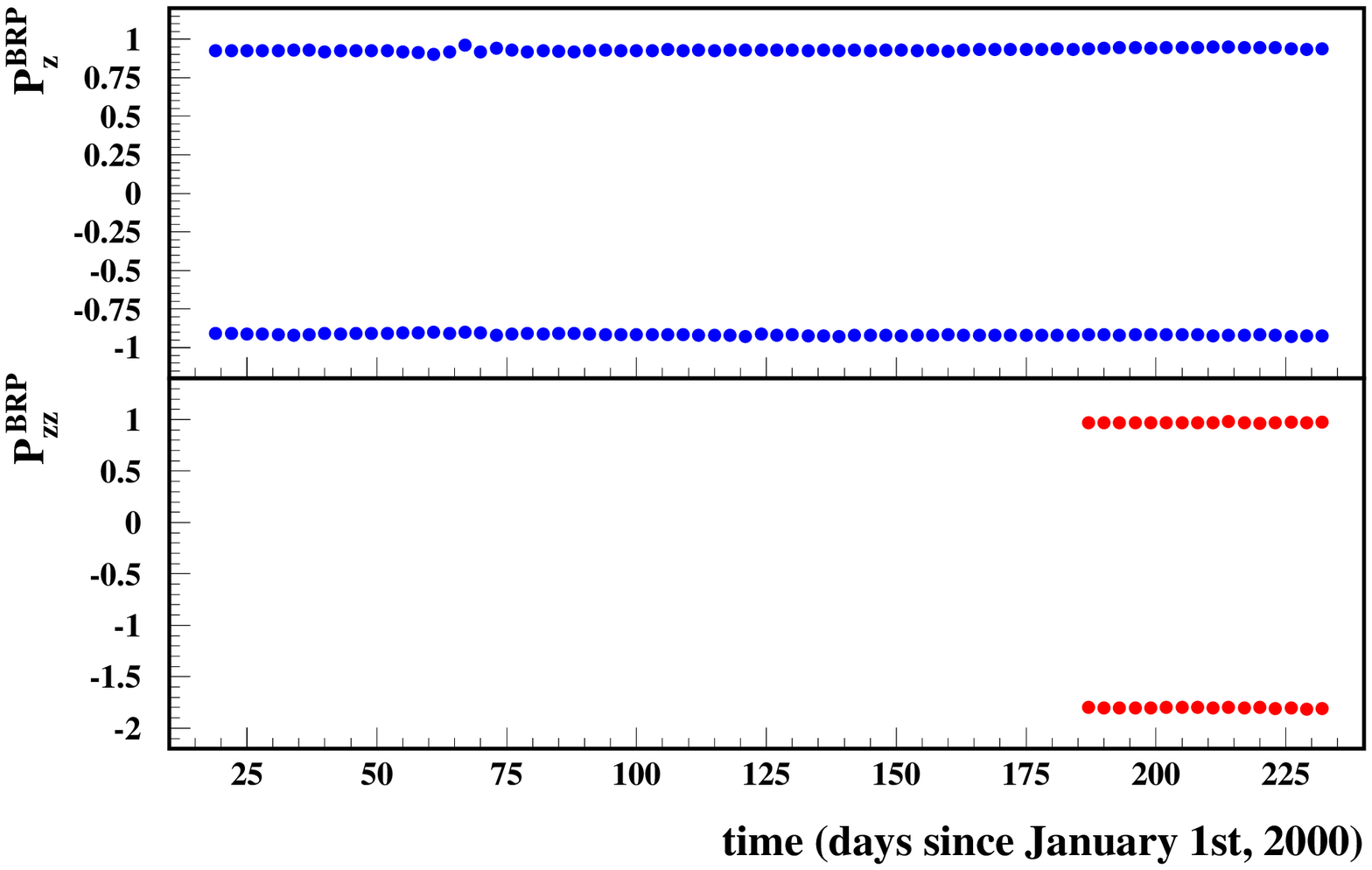} 
\end{tabular}
\end{center}
\caption{Atomic fraction, $\alpha^{TGA}$ upper plot, and vector/tensor
polarization, $P^{BRP}$ lower plots, measured by TGA and BRP respectively
during the year 2000 running period. Each symbol represents data
averaged over a 72-hour bin. The absolute bin averaged values 
are always above 0.9. The tensor polarization (lower plot) was employed 
from July on.}
\label{fig:raw2}
\end{figure*}

The outstanding performance of the HERMES target in this period is well 
represented in Fig.~\ref{fig:raw2} (note the change in scale compared to 
Fig.~\ref{fig:rawmeas}). 
The stable behavior of the cell surface in terms of atomic recombination 
is demonstrated in the upper plot of Fig.~\ref{fig:raw2}, where the atomic 
fraction measured by the target gas analyzer is plotted.
The lower plot shows the nuclear 
polarization. In the period between January and 
June 2000, the target was operated with only positive and negative states of
nuclear vector polarization, implying positive tensor polarization. 
From July on, a new injection mode combining 
vector and tensor polarization $P_{z+}$, $P_{z-}$, $P_{zz+}$ and 
$P_{zz-}$ was established in order to perform the first measurement of 
the tensor-polarized structure function $b_1^d$.

The average values for the four polarization states used in 2000 are the
following:
\begin{eqnarray}
P_{z+}&=&+0.851\pm0.029\\
P_{z-}&=&-0.840\pm0.026\\
P_{zz+}&=&+0.891\pm0.027\\
P_{zz-}&=&-1.656\pm0.049\,.
\end{eqnarray}
Note that only the systematic uncertainty is quoted as the statistical 
uncertainty is negligible in comparison.
The differences between the systematic uncertainties 
originate from the injection mode dependent systematic uncertainty of the 
atomic polarizations.
The measured target density for this period, measured by means of the
spin-exchange collision relaxation method, is 
$2.1\times10^{14}$~nucl./cm$^2$.
%%%%%%%%%%%%%%%%%%%%%%%%%%%%%%%%%%%%%%%%%%%%%%%%%%%%%%%%%%%%%%%%%

%%%%%%%%%%%%%%%%%%%%%%%%%%%%%%%%%%%%%%%%%%%%%%%%%%%%%%%%%%%%

\subsubsection{Spin relaxation}

\begin{figure*}
\begin{center}
\includegraphics[width=\textwidth]{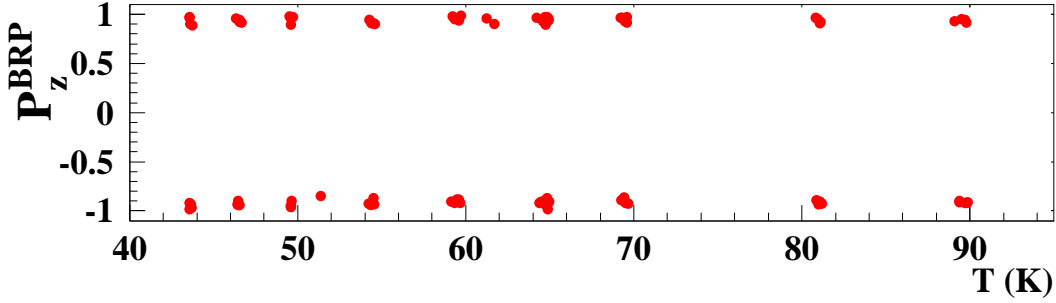} 
\end{center}
\caption{Nuclear vector polarization $P_z^{BRP}$ as function of the cell temperature
T measured in August 2000. No dependence on the temperature is observed.
The operating temperature of the target during data collection
was 60\,K.}
\label{fig:pzt}
\end{figure*}

The spin-flip probability per wall collision $\gamma_d$ is proportional
to $\tau_s$, the mean residence time:
$\tau_s=\tau_0\,e^{\bigl(\frac{E_b}{kT}\bigr)}$ (see eq.~(\ref{eq:arrenius})).
 Therefore 
$\gamma_d$ depends on the wall temperature $T$.
In other words, if $P_a^{BRP}$ does not change with $T$,
wall depolarization effects are vanishingly small 
within the experimental uncertainties.
In Fig.~\ref{fig:pzt}, 
a scan of the cell temperature down to very low temperatures is
shown. The result indicates that no wall depolarization occurs.
%The occurrence of wall depolarization can be revealed by exploiting 
%its temperature dependence. Indeed, 
%the probability of nuclear spin relaxation $\gamma_z$ is proportional to 
%the square of the sticking time 
%$\tau_s=\tau_0\,e^{\bigl(\frac{E_b}{kT}\bigr)}$ of the atom on the cell wall, 
%where $T$ represents the temperature of the surface 
%(see eq.~\ref{eq:arrenius}). Hence, for a given temperature range 
%$\Delta T$, the occurrence of a non-zero probability $\gamma_z>0$ can 
%be demonstrated by the measurement of a variation $\Delta P_z$ of the 
%nuclear polarization. During the whole running period in 2000, no such 
%a dependence has ever been observed at any reachable temperature value, 
%as represented in figure \ref{fig:pzt}, showing the results of a scan 
%performed in August 2000. 
As a consequence, for the deuterium target 
the sampling correction for nuclear depolarization equals unity ($c_P=1$), 
so that:
\begin{equation}
P_a=P_a^{BRP}\,.
\end{equation}
Moreover, the calculated depolarization due to spin 
exchange collisions resulted in $\Delta P_a^{se}=0.3\,\%$. 
As in 2000 spin relaxation can be neglected completely, 
the polarization injected by the ABS equals the measured BRP value:
\begin{equation}
P_a^{inj}\simeq P_a^{BRP}\,.
\end{equation}
The polarizations measured by the BRP were found to be consistent
with those calculated as described in Sec.~\ref{sec:injectedpol},
within the uncertainties.

\subsubsection{Recombination}

\begin{figure*}
\begin{center}
\begin{tabular}{c}
\includegraphics[width=\textwidth]{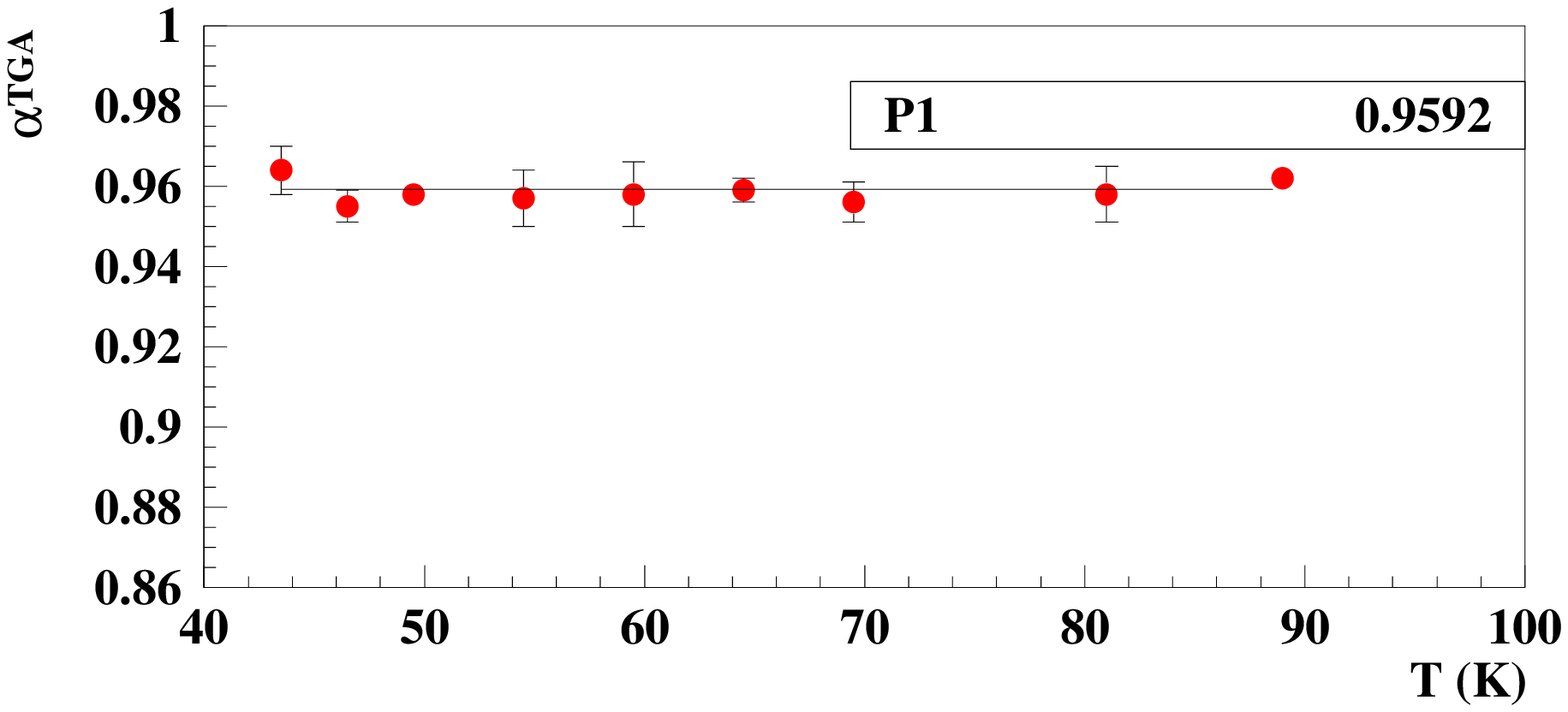}\\
\includegraphics[width=\textwidth]{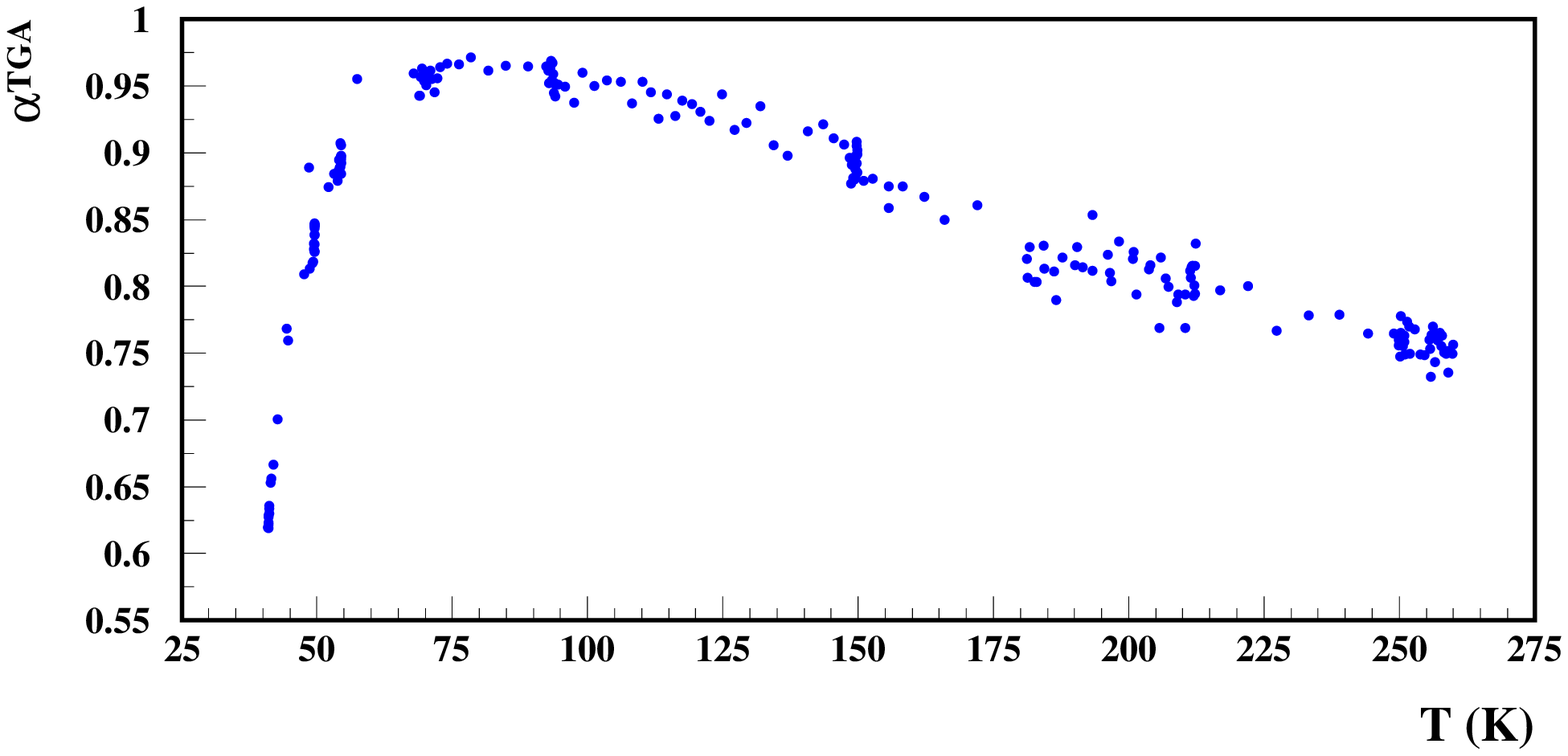} 
\end{tabular}
\end{center}
\caption{Upper plot: measured atomic fraction as a function of the cell 
temperature for $T_{cell}<100\,$K for the deuterium running in 2000. 
No dependence on the temperature can be seen. The working 
temperature during normal operation was set to $T_{cell}=60\,$K. 
For comparison a temperature scan with
hydrogen taken in 1997 is shown in the lower plot.}
\label{fig:tscan}
\end{figure*}

Although atomic recombination depends strongly on temperature, 
no variation of the measured atomic fraction 
$\alpha^{TGA}$ over a very wide temperature range 
around the working temperature $T_{cell}=60\,$K 
was ever observed during the whole year. Fig.~\ref{fig:tscan} (upper plot) 
reports a cell temperature scan performed in August. 
(For comparison a temperature scan performed with hydrogen in 1997 is 
also shown in the lower plot of the same figure).
Additionally the measurement of the fraction $\alpha_r^{TGA}$
of atoms surviving recombination yielded the following average value:
\begin{equation}
\alpha_r^{TGA}=0.997\pm 0.014\,,
\end{equation}
confirming that within the measured uncertainty no evidence 
of recombination was found.
Therefore, for the deuterium operating period in 2000, both the 
deviation from unity of the sampling correction $c_{\alpha}$, 
and the impact of the uncertainty in $\beta$ are negligible.

%%%%%%%%%%%%%%%%%%%%%%%%%%%%%%%%%%%%%%%%%%%%%%%%%%%%%%%%%%%%%%%%%%%

\subsection{Transverse hydrogen running in 2002}

%%%%%%%%%%%%%%%%%%%%%%%%%%%%%%%%%%%%%%%%%%%%%%%%%%%%%%%%%%%%%%%%%%

During the 2002-2003 operating period, the HERMES transversely polarized 
hydrogen target worked in a very stable way. On the other hand, 
the poor performance of the HERA beam did not allow a detailed study of 
the bunch induced depolarization. Nevertheless, by scanning the target 
magnetic field a working field value was established where the loss of 
polarization due to the field non-uniformity was limited to 1.5\,\% 
at the highest available beam currents (25 mA).
The average target polarization $(P_{z+} - P_{z-})/2$
during the 2002 data taking period was $0.783\pm 0.041$.
%This value is lower than that reported for the longitudinal hydrogen target 
%(Section 6.1), mainly due to the lower applied target holding field.
This value is lower than that reported for the longitudinal hydrogen target 
(Section 6.1), mainly due to the lower applied target holding field and to the
fact that the higher density allowed by the use of the intermediate size cell 
resulted in an increase of spin relaxation due to spin exchange collisions.

\begin{figure*}
\begin{center}
\begin{tabular}{c}
\includegraphics[width=\textwidth]{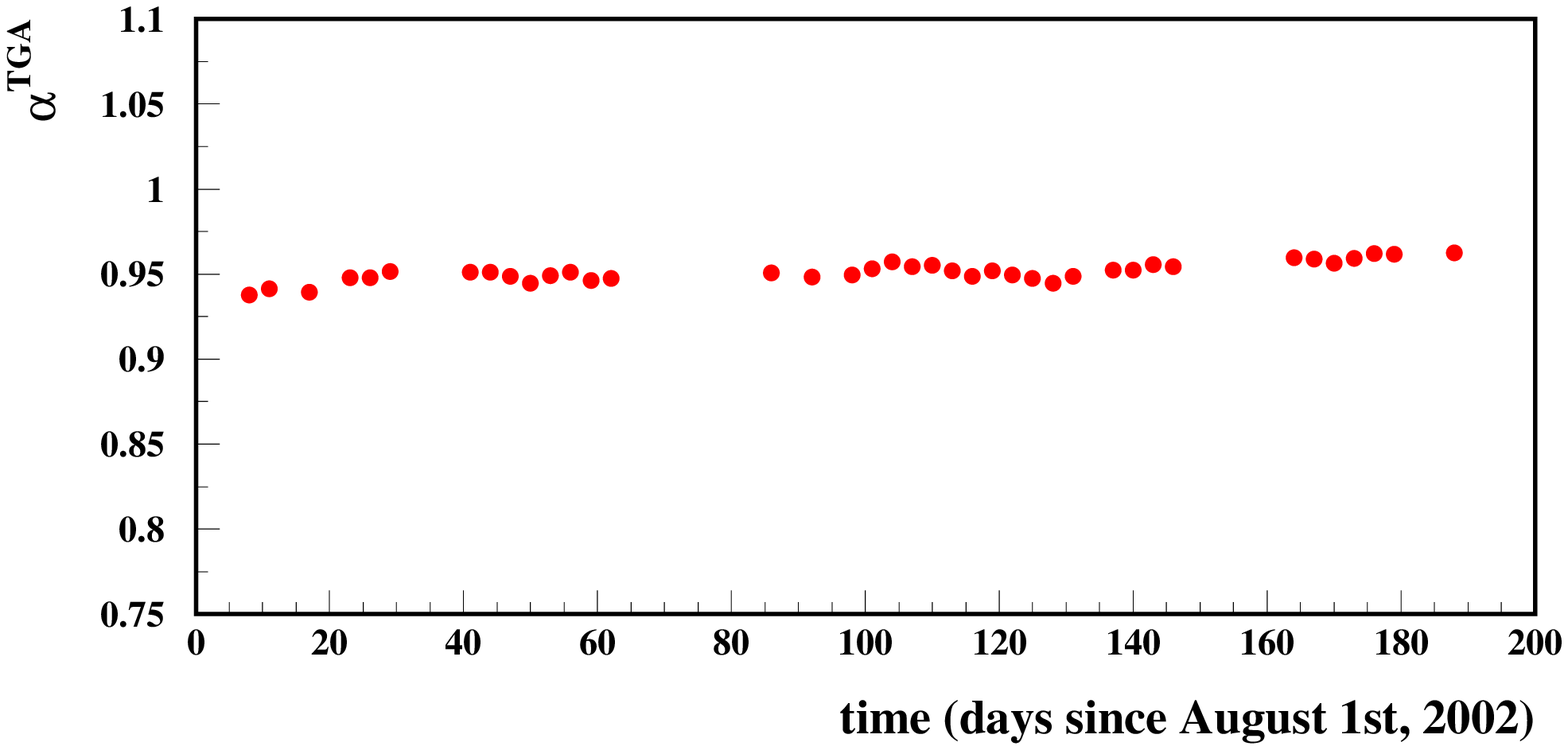} \\
\includegraphics[width=\textwidth]{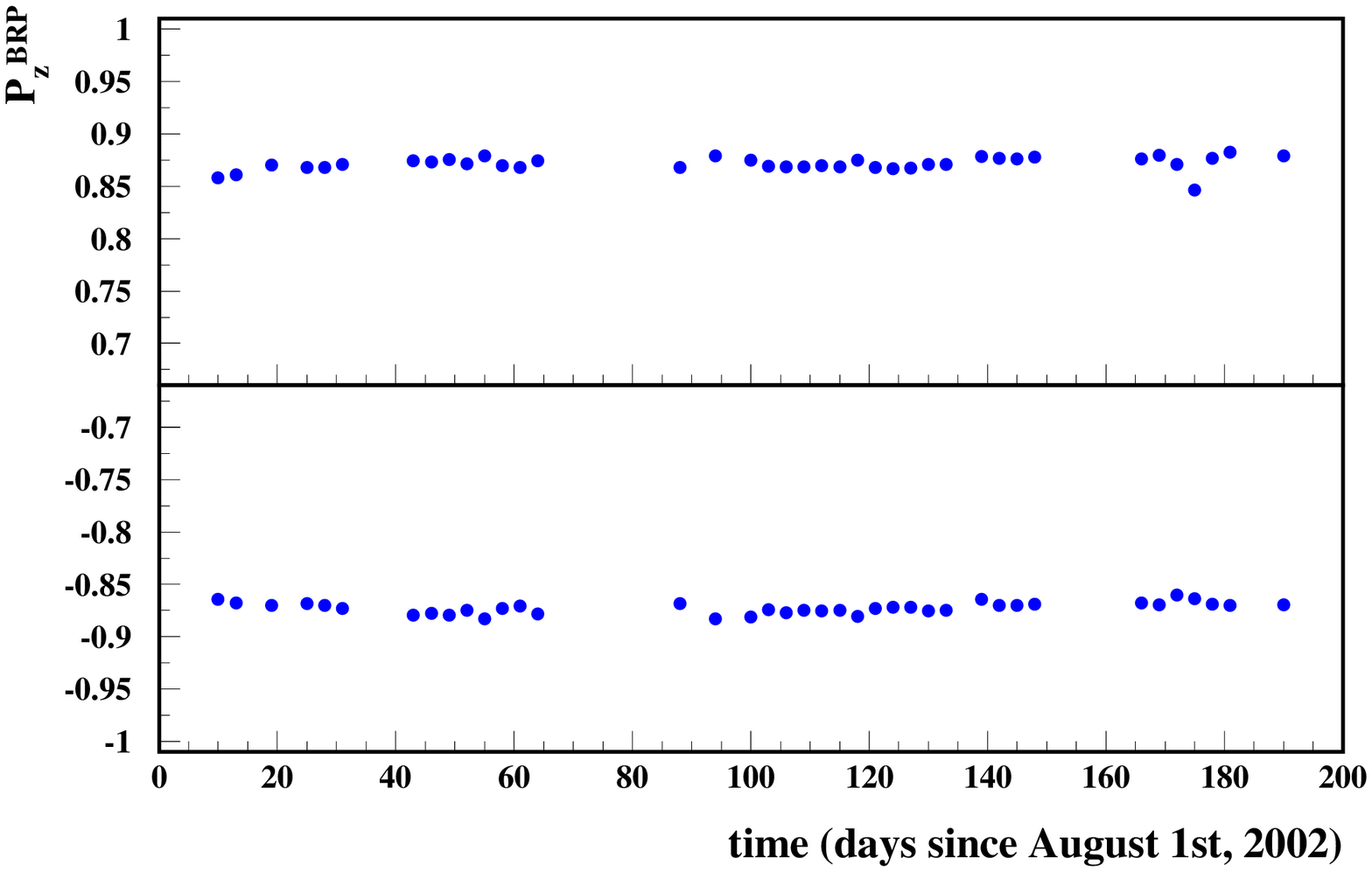} 
\end{tabular}
\end{center}
\caption{The TGA (above) and BRP (below) measurements for the
2002/03 data taking period.}
\label{fig:raw3}
\end{figure*}

The stability of the transverse 
target operation during the 2002 running period is illustrated in 
Fig.~\ref{fig:raw3}.
The target density in this period measured via spin-exchange
collisions was $1.1\times10^{14}$ nucl./cm$^2$.
%%%%%%%%%%%%%%%%%%%%%%%%%%%%%%%%%%%%%%%%%%%%%%%%%%%%%%%%%%%%%%%%%%%%%%

%%%%%%%%%%%%%%%%%%%%%%%%%%%%%%%%%%%%%%%%%%%%%%%%%%%%%%%%%%%%%%%%%%%%%%%

%%%%%%%%%%%%%%%%%%%%%%%%%%%%%%%%%%%%%%%%%%%%%%%%%%%%%%%%%%%%%%%%%%%%%
%\input{concl.tex}
\section{Discussion}

Tab.~\ref{poltable} summarizes averaged values for the target properties
during the various running periods.
%%%%%%%%%%%%%%%%%%%%%%%%%%%%%%%%%%%%%%%%%%%%%%%%%%%%%%%%%%%%%%%%%%%%%
\begin{table*}
\begin{center}
\begin{tabular}{|c|c|c|c|}
\hline
 & $H_{||} (1997)$ &$H_{\perp} (2002)$ &  $D_{||} (2000)$ \\ \hline\hline
$\alpha_0$ & $0.960\pm0.010$ & $0.918\pm0.032$ & $0.919\pm0.026$ \\ \hline
$\alpha_r$ & $0.945\pm0.035$ & $0.979\pm0.023$ & $0.997\pm0.017$ \\ \hline
$P_{z+}$ & $+0.908\pm0.016$ & $+0.859\pm0.032$ & $+0.927\pm0.017$ \\ \hline
$P_{z-}$ & $-0.908\pm0.016$ & $-0.859\pm0.032$ & $-0.915\pm0.010$ \\ \hline
$-\Delta P_{SE}$ & 0.035 & 0.055 & $\le 0.001$ \\ \hline
$-\Delta P_{WD}$ & 0.02 & 0.055 & $\le 0.01$ \\ \hline
$-\Delta P_{BI}$ & absent & 0.015 & absent \\ \hline
$P_{+}$ & $+0.851\pm0.031$ & $+0.783\pm0.041$ & $+0.851\pm0.029$ \\ \hline
$P_{-}$ & $-0.851\pm0.031$ & $-0.783\pm0.041$ & $-0.840\pm0.026$ \\ \hline
$t$ ($\mathrm{10^{14}nucleons/cm^{2}}$) &  0.7 & 1.1 & 2.1 \\ \hline
FOM ($P^2 t$) ($\mathrm{10^{14}/cm^2}$) & 0.5 & 0.67 & 1.5 \\ \hline
\end{tabular}
\end{center}
\caption{Summary of the HERMES target performance in the years 1997, 2000 and
2002. The tabulated values are averaged over the respective running periods.
The $\Delta P_{i}$'s represent the polarization losses arising from 
the listed process (see text).}
\label{poltable}
\end{table*}
%%%%%%%%%%%%%%%%%%%%%%%%%%%%%%%%%%%%%%%%%%%%%%%%%%%%%%%%%%%%%%
A comparison between the longitudinally polarized hydrogen and 
deuterium target leads to the following conclusions.
At the same guide field, in the case of deuterium 
the spin exchange and wall relaxation processes are suppressed 
by a factor $(B_C^H/B_C^D)^2\simeq 20$ compared to hydrogen.
The positive and negative $P_z$ values for hydrogen coincide very well,
while for deuterium there is a clear difference. This is due to the 
larger number of efficiencies and transmissions involved.

From the table it is possible to deduce departures from
the optimal performance like the amount of recombination 
($\Delta \alpha_r$ = $1-\alpha_r$), and 
the depolarization due to the various
depolarizing mechansims: spin-exchange collisions ($\Delta P_{SE}$), 
wall-depolarization 
($\Delta P_{WD}$) and beam-induced fields ($\Delta P_{BI}$). The target
density integrated over the cell length, or areal density $t$, is also
shown together with a relative factor of merit (FOM) that takes 
into account both the polarization and the density.
It is worthwhile to mention again that in 1997 the 
target cell had a larger cross-section (29\,mm$ \times 9.8\,$mm) implying 
a higher conductance and a correspondingly lower density (about 40\,\%).
The table also illustrates once more the optimal performance of the $D_{||}$
target in 2000. For this target no recombination ($\Delta\alpha_r=0$) 
and no depolarization phenomena ($\Delta P_{WD} = \Delta P_{SE} = 
\Delta P_{BI} = 0$) have been observed.
This excellent behavior has to be attributed to the lower critical field
of deuterium (11.7 mT) compared to hydrogen (50.7 mT), and to an
improvement in the knowledge and preparation of the cell
surface coating.
The improved surface coating is also responsible for the better 
performance of the 2002 $H_{\perp}$ target compared to the 1997 $H_{||}$ 
target, as far as the amount of recombination is concerned.

On the other hand,
in 2002, due the higher density obtained and the higher number of wall
collisions, spin-relaxation effects became more important, causing
 a loss in polarization of about 11\,\%
and limiting the gain in the factor of merit with respect to the 1997 
running due to the increase in density.
This indicates how an enhancement in the figure of merit of the target
might be obtained by
increasing the holding field proportionally with the density, 
in order to limit the spin relaxation effects.

\section{Selected HERMES results}
In this last Section before the conclusions, some
of the measurements which have been carried out at HERMES with the
polarized internal gaseous target during the various running periods are
mentioned.
The number of deeply inelastic scattering
events (after data quality selection) collected with the three polarized
targets is given in Tab.~\ref{table:DIS}.
%A maximum number of 8 million events was collected in the run 2000.
\begin{table}
\begin{center}
\begin{tabular}{|c|c|}
\hline
target (year) & $DIS/million$ \\ \hline \hline

$H_{||} (1996-97)$ & $2.4$ \\ \hline

$D_{||} (1998-2000)$ & $9.1$ \\ \hline

$H_{\perp} (2002-03)$ & $0.7$ \\ \hline

\end{tabular}
\end{center}
\caption{Deeply inelastic scattering events collected
with the three polarized targets after data quality selection. 
The numbers are given in millions.}
\label{table:DIS}
\end{table}

Among the measurements which have been performed using the HERMES
detector, we quote the following few: 
the measurement of the $g_1^p$ structure function~\cite{hermes:g1p}; the
measurement of the $g_1^d$ structure function~\cite{hermes:g1d}; 
the first
measurement of the $b_1^d$ structure function~\cite{hermes:b1d}; 
measurements of the quark-flavour decomposition of the nucleon 
spin~\cite{hermes:deltaq1,hermes:deltaq2}; the
first measurements of single-spin asymmetries in deep-inelastic
scattering off both longitudinal and 
transversely polarized targets~\cite{hermes:salongh,hermes:salongd,hermes:h1} 
and a first study of the
gluon polarization in nucleons~\cite{hermes:deltag}.

\section{Conclusion}
The HERMES polarized hydrogen/deuterium target has been reliably 
operating since 1996. It provides polarization values between 0.78 and 0.85,
densities 1-2 $\times$ $10^{14}$ nucleons/$\mathrm{cm^2}$, and total systematic 
uncertainties of only 0.03-0.04.
The systematics of both the recombination and depolarization processes in the
storage cell are fully under control.
%The results of the last running with hydrogen and increased target density
%evidenced the necessity of increasing the present holding field of about
%300 mT to further improve the factor of merit of the targets.

\section{Acknowledgements}
We gratefully acknowledge the DESY management for its
support and the staff at DESY and the collaborating institutions for
their significant effort. This work was supported by the FWO-Flanders,
Belgium; the Natural Sciences and Engineering Research Council of
Canada; the National Natural Science Foundation of China; the INTAS
and ESOP network contributions from the European Community; the German
Bundesministerium f\"ur Bildung und Forschung; the Deutsche
Forschungsgemeinschaft (DFG); the Deutscher Akademischer
Austauschdienst (DAAD); the Italian Istituto Nazionale di Fisica
Nucleare (INFN); Monbusho International Scientific Research Program,
JSPS, and Toray Science Foundation of Japan; the Dutch Foundation for
Fundamenteel Onderzoek der Materie (FOM); the U. K. Engineering and
Physical Sciences Research Council and the Particle Physics and
Astronomy Research Council; and the U. S. Department of Energy and the
National Science Foundation.

%\bibliographystyle{elsart-num}
%\bibliography{ref}

\end{document}